\documentclass[draft]{agujournal2019}
\usepackage{url} 
\usepackage{lineno}
\usepackage{stmaryrd}
\usepackage[inline]{trackchanges} 
\usepackage{soul}
\usepackage{textcomp}
\usepackage[T1]{fontenc}

\usepackage{commath}
\usepackage{todonotes}

\usepackage[utf8]{inputenc}
\DeclareUnicodeCharacter{2202}{-}
\DeclareUnicodeCharacter{221A}{-}

\newcommand{\iu}{{i\mkern1mu}}

\draftfalse

\journalname{JAMES}

\begin{document}

%
%

\title{Assessing the Potential of Deep Learning for \\Emulating
Cloud Superparameterization in \\
Climate Models with Real-Geography Boundary Conditions}



%
%




\authors{Griffin Mooers\affil{1,2}, Michael Pritchard\affil{1}, Tom Beucler\affil{1}, Jordan Ott\affil{2,1}, Galen Yacalis\affil{4}, Pierre Baldi\affil{2}, Pierre Gentine\affil{3}}

\affiliation{1}{Department of Earth System Science, University of California at Irvine, CA, USA}
\affiliation{2}{Department of Computer Science, University of California at Irvine, CA, USA}
\affiliation{3}{Department of Earth and Environmental Engineering, Columbia University, New York, NY, USA}
\affiliation{4}{Department of Mathematics, University of California at Irvine, CA, USA}





\correspondingauthor{Griffin Mooers}{gmooers96@gmail.com}




\begin{keypoints}

\item Feed-forward neural networks can emulate explicit convection including geographic complexity.
\item After tuning, our neural network can fit 70 percent of the temporal variance in the mid-to-upper troposphere.
\item Deep convection over land is parameterizable in neural networks locally in time.


\end{keypoints}

%
%

%
%


\begin{abstract}

We explore the potential of feed-forward deep neural networks (DNNs) for emulating cloud superparameterization in realistic geography, using offline fits to data from the Super Parameterized Community Atmospheric Model. To identify the network architecture of greatest skill, we formally optimize hyperparameters using $\sim$250 trials. Our DNN explains over 70 percent of the temporal variance at the 15-minute sampling scale throughout the mid-to-upper troposphere. Autocorrelation timescale analysis compared against DNN skill suggests the less good fit in the tropical, marine boundary layer is driven by neural network difficulty emulating fast, stochastic signals in convection. However, spectral analysis in the temporal domain indicates skillful emulation of signals on diurnal to synoptic scales. A close look at the diurnal cycle reveals correct emulation of land-sea contrasts and vertical structure in the heating and moistening fields, but some distortion of precipitation. Sensitivity tests targeting precipitation skill reveal complementary effects of adding positive constraints vs. hyperparameter tuning, motivating the use of both in the future. A first attempt to force an offline land model with DNN emulated atmospheric fields produces reassuring results further supporting neural network emulation viability in real-geography settings. Overall, the fit skill is competitive with recent attempts by sophisticated Residual and Convolutional Neural Network architectures trained on added information, including memory of past states. Our results confirm the parameterizability of superparameterized convection with continents through machine learning and we highlight advantages of casting this problem locally in space and time for accurate emulation and hopefully quick implementation of hybrid climate models.

\end{abstract}


\section*{Plain Language Summary}

Machine learning methods have been previously used to replace parameterizations (approximations) of atmospheric convection under very idealized scenarios (aqua-planets). The hope is that these machine learning emulators can help power a next generation of climate models with similar accuracy but at a fraction of the computational cost. But important questions remain about how learnable more realistic convection (over both land and ocean) is. Recently, a first attempt at machine learning replicated convection was made under these Earth-like conditions. But it required a highly specialized neural network as well as memory of the previous behavior of the atmosphere. This design would make using these machine learning emulators with climate models very difficult. This motivates learning convection under realistic geography with a simpler network. Our results are reassuring because our simple neural network learns realistic convection over land as well as a more complicated model. But even harder tests involving full coupling with a host climate model will be needed to truly test this method's potential.

%
%

%


%
%
%
%

\section{Introduction}
\par
Although global atmospheric model simulations are increasingly high-resolution, even under optimistic scenarios of enhanced computing performance, physically resolving the atmospheric turbulence controlling clouds will likely not be feasible for decades. Current climate model horizontal grid cells are typically 50-100 kilometers wide but the turbulent updrafts governing cloud formation occur on scales of just tens to hundreds of meters and the microphysical processes regulating convection occur down at the micro-meter scale~\cite{Schnieder_2017, LES_Comp, Morrison_2020}. This discrepancy creates large uncertainties about the precise details of deep convection on cloud feedbacks and climate change~\cite{Sherwood_Nature}. Multi-scale methods such as embedding two-dimensional Cloud Resolving Models (CRMs) into General Circulation Model (GCM) grid cells (superpararmeterization) have been used to directly resolve the spatial and temporal progression of moist convection. More recently, explicit kilometer-scale simulation of moist convection has improved the representation of deep convective clouds and the hydrological cycle~\cite{10.1175/BAMS-84-11-1547, Schnieder_2017, Li_Xie,Christensen, Daleu_2015}. These advancements allow models to simulate historically challenging atmospheric modes of variability like the observed afternoon maxima of deep convection over continents and a more realistic probability distribution of precipitation that captures extremes on the tail-end~\cite{kooperman_2016, Li_et}. However, even the highest resolution global CRM simulations today require some assumption-prone parameterization for microphysics and sub-km turbulence, among other cloud processes~\cite{Siebesma_2007, Cheng_Xu}, although multi-scale algorithms still hold promise for making some of this explicitly tractable~\cite{Parishani_Pritchard, janssonetal2019}.

\par
Given these dual physical and computational hurdles, using machine learning emulators to replace sub-grid convective physics in coarse-resolution climate models is an area of rapidly increasing interest. Following seminal works including~\citeA{Krasnopolsky_2010, KRASNOPOLSKY2008535, Krasnopolsky_2013}, recent breakthroughs from global aqua-planet simulations have provided a proof of concept for hybrid climate models powered by machine learning.~\citeA{Gentine_all} showed 40-100M samples taken from a zonally symmetric aqua-planet simulation were sufficient to train a five to ten layer DNN to emulate superparameterized convective heating and moistening in a hold-out test dataset, with $R^{2}$ greater than 0.7 in the midtroposphere. Building on these results,~\citeA{Rasp9684} demonstrated that a similar DNN could even be run in a prognostic setting, coupled to an advective scheme in the Community Atmosphere Model (CAM), thus generating accurate mean climate states and equatorial wave spectra at as low as five percent of the computational cost of actual superparameterization. Recently,~\citeA{OGorman_Dwyer, Yuval_Ogorman} showed that similar prognostic success can occur in idealized aqua-planets trained on coarse-grained three-dimensional output using Random Forests (RFs). These RFs used refined inputs and outputs tailored to the prognostic variables underlying the System for Atmospheric Modeling or SAM, which is the embedded cloud resolving model used in the SPCAM multimodel framework. Whereas all of the above studies have focused on aqua-planets, skillfully replicating convection in more complex, realistic settings is a key step towards building a replacement for traditional sub-grid parameterizations of deep convection in climate models.


\par

Achieving competitive emulation of convection under realistic geography may be a much more significant hurdle for neural networks. At the time of this writing, a first pioneering attempt has been made to fit superparameterized convection in a realistic operational setting. Results of this study indicate that sophisticated network designs involving the addition of 1D convolutions in the vertical dimension, and `residual'' neural network architecture~\cite{he2015deep} using state information from previous time steps, appeared critical to achieving reasonable fits~\cite{hanetal2020}. This raises two issues. From a practical perspective, implementing neural networks that rely on prior temporal information (such as the Resnet in~\citeA{hanetal2020}) as coupled components of a host climate model is technically challenging since previous timesteps are not typically passed to the physics parameterization. From a philosophical perspective, this questions casting machine learning parameterizations of convection locally in time. If confirmed, these two issues makes the full potential of neural network (NN) emulators as a tool to advance scientific understanding beyond aqua-planets substantially harder to utilize.

On this basis, we explore whether feed-forward DNNs are capable of emulating convection with real-geography if sufficient hyperparameter tuning is taken advantage of when training our neural network. The hypothesis that even a feed-forward neural network can emulate superparameterized convection with realistic geography is based in part off the results of~\citeA{ott2020fortran} on aqua-planets in which formal, expansive hyperparameter tuning was identified as essential for the performance of a DNN to emulate convection. We focus on emulating superparameterization to avoid the ambiguities due to coarse-grained uniformly-resolved CRM output~\cite{Brenowitz_Bretherton}. We readily acknowledge that other methods such as RFs~\cite{yuval2020use} also show success and promise in sub-grid convection emulation. Indeed, RFs have some advantages, including automatically respecting physical constraints that are linear in their outputs, such as energy conservation, and positive-definite precipitation, which are not guaranteed in DNNs~\cite{yuval2020use}. Furthermore,~\citeA{Meyer_2021} has demonstrated that RFs can be used to correct, or "nudge" parameterizations and reduce simulation errors even for realistic convection. However, there are also ways to enforce such constraints in DNNs~\cite{beucler2019enforcing, ReLU_fix}, and RFs also come with disadvantages. To cite a few, RFs with deep trees quickly become computationally expensive for large datasets, requiring large storage capacity which could prevent taking full advantage of Graphics Processing Unit (GPU) infrastructure ~\cite{yuval2020use}. RFs may struggle to capture local patterns in the atmosphere as well~\cite{Meyer_2021}. For these reasons, we leave RFs for future work.

Here, our task is to understand what convective patterns, cycles, and modes of variation in a realistic setup of superparameterized convection can be fitted with a feed-forward DNN. We additionally aim to establish a set of post-processing metrics to benchmark our own neural network's performance and transparently compare different neural network emulators trained on similar data. Section~\ref{Methods} outlines the details of our simulation dataset, introduces the design of a neural network, and describes our automated hyperparameter tuning algorithm, capable of finding a reasonable fit. Then, in Section~\ref{METRICS}, we lay out our test benchmarks. In Section~\ref{Results} we present the spatial and temporal breakdown of our neural network predictions for parameterized convective tendencies. We also analyse the plausibility of the neural network emulated hydrological cycle in detail. The last part of Sections~\ref{Results} examines the potential to couple an aqua-planet trained neural network to a land model as another credibility test towards a hybrid climate model. Section~\ref{sec:Conclusion} includes a summary of our work, its limitations, and potential directions for future research.

\section{Methods}\label{Methods}

\subsection{Climate Simulation Data}\label{Simulation}
We leverage three different datasets to train, test, and benchmark DNNs emulating convection with real-geography. The data are based on the Super Parameterized Community Atmospheric Model (SPCAM), a global climate model that nearly explicitly resolves atmospheric moist convection by using idealized embedded CRMs~\cite{10.1175/BAMS-84-11-1547, Grabowski_2001}. Each of the host GCM's grid cells embeds two-dimensional CRMs of optional horizontal resolution and physical extent, thus avoiding heuristic parameterization of sub-grid moist convective processes~\cite{10.1175/BAMS-84-11-1547, Benedict_Randall}.

\par
For a point of reference, we first use outputs from SPCAM v.3 (SPCAM3) at T42 spectral truncation (i.e. 8,192 horizontal grid cells) driven with boundary conditions of a zonally symmetric aqua-planet; as in~\citeA{Rasp9684}. We then build beyond previous aqua-planet emulation studies by generating a new dataset from a more modern version (v.5) of SPCAM (SPCAM5) that includes higher horizontal resolution (1.9x2.5 degree finite volume dynamical core, i.e. 13,824 grid cells) and in which we incorporate realistic boundary conditions, including a land surface model, seasonality, and a zonally asymmetric annual climatology of sea surface temperatures (SSTs). The dataset is similar to one recently used in~\citeA{hanetal2020} but with a few differences. The simulation itself is 10 years long, but selectively sub-sampled to every 10 days to avoid temporally autocorrelated training samples. We also rely on a shorter GCM timestep (15 minutes as opposed to 20) (Table~\ref{tab:SPCAM_Table}). As in~\citeA{Gentine_all,Rasp9684}, but unlike~\citeA{hanetal2020}, we make the further simplifiation of using a reduced (32-km) CRM horizontal extent, i.e. CRMs with only 8-columns apiece, instead of the 128-km / 32-column CRM configuration (Table~\ref{tab:SPCAM_Table}). This decision, based on~\citeA{pritchetal2014}'s finding that (for deep convection) small CRM domains do not corrupt the representation of tropical wave dynamics in SPCAM. This also has the advantage of simplifying the comparison of our results to~\citeA{Gentine_all,Rasp9684}. Meanwhile there are reasons to think it may facilitate DNN emulation~\cite{brenowitz2020interpreting,ott2020fortran}. The codebase for running the ``SPCAM5'' simulations is the same employed by~\citeA{Parishani_Pritchard}, which is archived at~\url{https://github.com/mspritch/UltraCAM-spcam2_0_cesm1_1_1}; this code was in turn forked from a development version of the CESM1.1.1 located on the NCAR central subversion repository under tag~\url{spcam_cam5_2_00_forCESM1_1_1Rel_V09}, which dates to February 25, 2013. 
Finally, for reference, we analyze output from the conventionally parameterized version of CAM5; this helps assess the emulation of Superparameterization compared to conventional parameterization.

\begin{table}
\centering
\begin{tabular}{ |p{4cm}||p{2.5cm}|p{2.5cm}|p{2.5cm}| }
 \hline
 \multicolumn{4}{|c|}{Simulation Datasets} \\
 \hline
 Details & CAM5 & SPCAM3 & SPCAM5\\
 \hline
 Spatio-temporal resolution& 1.9\textdegree $\times $2.5\textdegree $\times $15min & 2.8\textdegree $\times $2.8\textdegree$\times $30min & 1.9\textdegree $\times $2.5\textdegree $\times $15min\\
 Total Number of Days Simulated & 93 & 93 & 3,650\\
 Total Number of atmospheric columns Simulated &123,420,672 &36,569,088 & 4,843,929,600\\
 \hline
\end{tabular}
\caption{Details of the three datasets used for benchmarking the results of our DNN trained on real-geography data.}
 \label{tab:SPCAM_Table}
\end{table}
\par

\subsection{Neural Network Design}\label{DNN}

\begin{table}
\centering
\begin{tabular}{ |c|c|c|c|c| }
\hline
\multicolumn{5}{|c|}{DNN Setup} \\
\hline
Input & Size & Output & Scaling factor & size \\
\hline
Temperature $(K)$ & 30 & Heating Tendency $(K / s)$ & $c_p$ & 30 \\
Specific Humidity $(kg / kg)$ & 30 & Moistening Tendency $(kg / kg / s)$ & $L_s$ & 30\\
Surface Pressure $(hPa)$ & 1 & TOA LW Flux $(W / m^2)$ & -1e-3 & 1\\
Solar Insolation $(W / m^2)$ & 1 &Surface LW Flux $(W / m^2)$ & 1e-3 & 1 \\
Sensible Heat Flux $(W / m^2)$ & 1 &TOA SW Flux $(W / m^2)$ & -1e-3 & 1\\
Latent Heat Flux $(W / m^2)$ & 1 & Surface SW Flux $(W / m^2)$ & 1e-3 & 1\\
 & & Precipitation $(m / s)$ & 1728000 & 1 \\
\hline
\end{tabular}
\caption{Details of the input and output vectors to the DNN. $c_{p}$ refers to the specific heat capacity of air at a constant pressure and is assumed to be 1.00464e3 $(J/kg/K)$ and $L_{s}$ is the latent heat of sublimination of water in standard atmospheric conditions calculated by adding the latent heat of vaporization 2.501e6 (J/kg) and the latent heat of freezing 3.337e5 (J/kg). Precipitation is weighted by the same prefactor, 1728000, also used in~\citeA{Rasp9684} to ensure it is felt in the loss function of the DNN.}
\label{tab:NN_Table}
\end{table}
We design a DNN that takes the same inputs as standard convection parameterizations in CAM to predict sub-grid scale tendencies at each vertical level and across each timestep globally. The neural network inputs can be thought of as atmospheric thermodynamics components in the eight year SPCAM5 data training simulation including: both temperature $(K)$ and specific humidity $(kg / kg)$ for each of the 30 vertical levels spanning the column, as well as surface latent heat flux $(W / m^{2})$, surface sensible heat flux $(W / m^{2})$, top of atmosphere (TOA) solar insolation $(W / m^{2})$, and surface pressure $(hPa)$. By including surface pressure in the input vector, we allow the neural network to fit horizontal variations in the vertical pressure grid, which is based on a hybrid terrain-following coordinate~\cite{Neale10descriptionof}. Concatenating these state variables creates an input vector to the neural network of length 64. Each of the input variables was pre-normalized before exposure to the neural network by subtracting its respective mean and dividing by its respective range, with these statistics computed and applied separately for each vertical level in the case of the vertically-resolved temperature and humidity profiles (Table~\ref{tab:NN_Table}). The reason we divide by the range instead of the more traditional standard deviation, in line with the methods of~\citeA{Rasp9684}, is to avoid dividing by near-zero numbers, e.g. in the case of stratospheric humidity. Some previous aqua-planet experiments also used the meridional wind vertical profile as part of the input vector to the neural network, but it was omitted in this case as preliminary neural network tests indicate it had an insignificant effect on the skill of the trained network while increasing the input vector length by 30 and thus substantially increasing training time~\cite{Rasp9684, Gentine_all}; we note that~\citeA{hanetal2020} also deem this an avoidable input. 

Our DNN ultimately predicts the sub-grid scale time tendency of temperature $(K / s)$ or heating tendency for short, which includes the sub-grid advection of temperature by convection and fine-scale turbulence, as well as grid average radiative heating throughout the column. It also predicts the sub-grid scale time-tendency of specific humidity throughout the column $(kg / kg / s)$ - or moistening tendency for short. A scalar is predicted for precipitation $(mm / day)$ as well as for the long and shortwave net radiative fluxes $(W / m^{2})$ at both the surface and the TOA. This fully concatenated output vector is of length 65 (Table~\ref{tab:NN_Table}). The state variables that comprise the output vector have different units, making the ultimate Mean Squared Error (MSE) of the neural network devoid of physically meaning. To ensure that all of the predicted variables have comparable magnitude and can be felt in the optimizer, we apply multiplicative prefactors as in~\citeA{Rasp9684}, recognizing that other choices can also be made such as additionally weighting by the mass of each pressure level~\cite{beucler2019enforcing}.

\subsection{Performance Analysis and Postprocessing}\label{METRICS}

To assess the skill of our DNNs after training, we benchmark them against an offline hold-out test dataset with multiple metrics. This is a first step to determine whether our DNNs could be candidates for online coupling, which we leave for future work. How well a neural network emulator appears to perform is in part a reflection of statistical analysis choices. Multiple conventions have been used and the degree of spatial averaging before applying error statistics has not been sufficiently reported to do inter-study comparison confidently, though spatial averaging is common practice~\cite{hanetal2020,Rasp9684, Brenowitz_Bretherton}. In some cases snapshots of unaveraged data~\cite{Rasp9684} have helped reveal issues at the finest resolved scales while zonally averaged temporal standard deviations~\cite{hanetal2020} have helped reveal issues in emulation of convective tendencies at small time intervals and spatial scales in neural network fits. Precipitation time series and Probability Density Functions (PDFs)~\cite{Rasp9684, hanetal2020, OGorman_Dwyer} have also been used to assess neural network performance.

In our case, to examine the magnitude of the error between the neural network prediction and the SPCAM5 target data, which we treat as truth, we will calculate a sum of squared errors (SSE) separately for each longitude and latitude and, in the case of 3D variables, vertical level (based on the hybrid, terrain following sigma coordinate):

\begin{linenomath*}
\begin{equation}
\mathrm{SSE \overset{\mathrm{def}}{=} \sum_{j=1}^{N_{t}}(y_{j}-\hat{y_{j}})^2}
\end{equation}
\end{linenomath*}
where $N_{t}$ is the length of the time series, $y$ is the target SPCAM5 data, and $\hat{y}$ is the corresponding neural network predicted value based on coarse-grained variables. In this case, we examine the performance of the neural network predicting heating and moistening tendencies. The primary metric for assessing DNN prediction and the associated spatial error structure is the coefficient of determination, ${R}^2$, defined as:

\begin{linenomath*}
\begin{equation}
\mathrm{{R}^2} \overset{\mathrm{def}}{=} 1-\frac{\mathrm{SSE}}{\mathrm{\sum_{j=1}^{N_{t}}(y_{j}-\bar{y}})^2},
\end{equation}
\end{linenomath*}
where $\bar{y}\ $is the temporally-averaged heating or moistening tendency at a given latitude, longitude, and vertical level. 

We apply ${R}^2$ to data entirely unaveraged in the latitude, longitude, and pressure. When visualizing averaged ${R}^2$, we often use a latitude-longitude cross sections at specific vertical levels to reveal error structure at the native 15 minute sampling interval (Figure~\ref{fig:R2_SFC}). In other portions of the analysis, we use spatial averaging prior to the error calculation, as in~\citeA{hanetal2020}. For instance, in pressure-latitude cross-sections (Figures~\ref{fig:SPCAM3},~\ref{fig:15min_vs_daily_zonalmean}), we first zonally average the predictions and targets, before computing ${R}^2$ over the time dimension. Furthermore, we examine $R^{2}$ at two different temporal resolutions: the native model timestep (15 minute sampling; Figures~\ref{fig:SPCAM3},~\ref{fig:R2_SFC},~\ref{fig:Autocorr}) since the strong diurnal cycle over land regions could bias the analysis between land and ocean regions, and then visualize with temporal averaging reducing the data to daily means (Figure~\ref{fig:15min_vs_daily_zonalmean}).

We also wish to elucidate whether there is any detectable ``mode-specific'' performance, i.e. certain temporal patterns that are especially predictable such as the diurnal cycle over continents in our moist convection emulation.
To that end, we calculate the temporal Power Spectral Density (PSD) for a single month (July):

\begin{linenomath*}
\begin{equation}\label{eq:PSD}
\mathrm{PSD_{k} \overset{\mathrm{def}}{=} \frac{2\Delta t}{N_{t}}\abs{\mathcal{F}\left(y\right)_k}^2},
\end{equation}
\end{linenomath*}
defined as the square complex modulus of the Fourier transform:

\begin{linenomath*}
\begin{equation}
\mathrm{ \mathcal{F}\left(y\right)_k \overset{\mathrm{def}}{=} \sum_{j=0}^{N_{t}-1} y_je^{\frac{ -2 \pi \iu jk}{N_{t}}}},
\end{equation}
\end{linenomath*}
where $\mathrm{y}$ is a time series of values of convective heating or moistening tendency at a given location, $\Delta t$ is the sampling time interval, and $\iu$ is the imaginary unit so that $\iu^{2}=-1$~\cite{1965-cooley}. The PSDs of heating and moistening tendencies are analysed both regionally and globally as follows: we mass weight each vertical level and then a PSD value is calculated for each frequency bin at each latitude, longitude, and pressure grid cell. We focus on timescales up to a month to examine variations in convection ranging from sub-diurnal to synoptic timescales. Next, the spectral coefficients at each latitude, longtiude, and pressure level are combined into a single, averaged PSD for the globe. We repeat this same analysis twice more, once with a land mask and once with an ocean mask. We also perform corresponding spectral analysis in the spatial domain i.e. calculating the PSD as a function of zonal and meridional wavenumbers, separately for every vertical level and model time-step over the same month of July. For the zonal spectrum, we restrict our average to just tropical locations from 20S to 20N and weight by the cosine of latitude to make an approximate Cartesian plane assumption. This enables us to sidestep the unequal grid spacing that would be a problem in this analysis if we included the mid-latitudes in our spatial average. 

Finally, to hone in on a regime in the lower tropical atmosphere that our ${R}^2$ analysis suggests is especially difficult for our DNN to emulate, we will analyze the temporal autocorrelations of the sub-grid scale tendencies. Our goal is to understand the regions where the DNN emulation of superparameterized convection is detectably worse than the global average performance. As a proxy for the ``stochasticity'' of atmospheric motions, we calculate the autocorrelation function (ACF) - a measure of the self similarity between a given signal and a delayed version of itself - using the previously calculated PSD:

\begin{linenomath*}
\begin{equation}
\mathrm{ACF_{j} \overset{\mathrm{def}}{=} \frac{1}{2\Delta t}\sum_{k=0}^{N_{t}-1}\left(\mathrm{PSD}_k\times\ e^{\frac{ 2 \pi \iu jk}{N_{t}}}\right)}
\end{equation}
\end{linenomath*}

 Fast signals that decorrelate quickly are more likely to be of stochastic nature. We thus compare the time to e-folding decay in $\mathrm{ACF}$ with ${R}^2$ skill score in the planetary boundary layer to test for correlations between DNN skill and the timescale of dominant atmospheric signals (diurnal cycle, Rossby waves) visible in vertically resolved heating and moistening tendencies (Equation~\ref{eq:PSD}). We also use the inverse of the e-folding decay timescale, which we will refer to as the ``autocorrelation frequency'', to examine the patterns between $R^{2}$ coefficient of determination globally and the stochasticity of the dominant convective signals. This comparison offers a possible explanation for much of the variations in the performance of our DNN throughout the planetary boundary layer.

To better quantify the differences between true and predicted PSDs, we rely on the Log-Spectral Distance (D)~\cite{Log_Dist}:

\begin{linenomath*}
\begin{equation}\label{eq:LSD}
\mathrm{D \overset{\mathrm{def}}{=} \sqrt{\frac{1}{N_{t}}\sum_{k=1}^{N_{t}-1}\left[\ln\left(\frac{PSD_{true_{k}}}{PSD_{pred_{k}}}\right)\right]^2}}
\end{equation}
\end{linenomath*}
where the summation is done in frequency space. We examine the neural network performance not just for the convective heating and moistening tendencies but also for precipitation, for which we both calculate the PDF and global error between DNN predicted precipitation and SPCAM5 target data. Furthermore, we determine the diurnal timing of maximum precipitation globally, but we seek to filter out noise in the mid-latitudes. To that end, when determining the hour of maximum precipitation rate, we look only at locations that pass the following threshold:

\begin{linenomath*}
\begin{equation}
\label{Precip_Sig}
\max\left(\mathrm{Precip}\right)-\min\left(\mathrm{Precip}\right)>\frac{2}{\sqrt{N_{\mathrm{day}}/4-1}}\max\left[\mathrm{std}\left(\mathrm{Precip}\right)\right]
\end{equation}
\end{linenomath*}
where $Precip$ refers to the model output precipitation rate in mm/day, $N_{day}$ refers to the number of days examined and the max refers to the local maximum of the precipitation in the temporal dimension at a given latitude, longitude grid cell. Our assumption is that the effective degrees of freedom in the diurnal composite is 1/4 of the apparent degrees of freedom. More empirically, if we relax the threshold anymore we detect unrealistic signals in the marine zones of the mid-latitudes.

\subsection{Formal Hyperparameter Tuning}\label{Tuning}

In several previous studies, small volumes of training data (as low as three months) and manual hyperparameter tuning were sufficient to achieve acceptable Machine Learning (ML) emulator performance~\cite{Rasp9684, Gentine_all}. Here, we make the hypothesis that with real-geography boundary conditions, neural networks benefit from considerably more training data and formal hyperparameter tuning~\cite{ott2020fortran}. To fully exploit our 10-year simulation, we split it into a training data set spanning the first eight years, a validation data set spanning the ninth year, and a test data set spanning the tenth year.

As a first step we subsampled by a factor of ten after sensitivity tests (not shown) indicated little difference in the fit skill from manual tuning attempts, likely due to redundant information from temporally autocorrelated state data. This subsampling in our preprocessing reduced the training data volume to a size which could be managed on a single GPU. Our initial architecture was inspired by previous literature, i.e. composed of five fully connected layers with 256 nodes each. However, this manual configuration yielded poor performance (Figure~\ref{fig:SPCAM3}), and other manual attempts to explore alternate choices of hyperparameters and learning rate variations were likewise unsuccessful (not shown). A higher fit skill is desirable before undertaking the difficult task of coupling the neural network to the host climate model for online analysis.

We attained much better results after adopting a formal hyperparameter tuning. Automated neural network architecture searches have just begun to prove their value in climate modeling - both for optimizing offline fits~\cite{beucler2019enforcing} and even prognostic online coupled performance~\cite{ott2020fortran}. Using similar approaches, we implemented a resource-intensive automated DNN training process, conducting a formal search over the following hyperparameters: batch normalization, dropout, LeakyReLU coefficient~\cite{Maas13rectifiernonlinearities}, learning rate, learning rate decay, number of layers, nodes per layer, and the optimizer~\cite{kingma2014adam}. All parameters and their corresponding ranges for the search are shown in Table~\ref{tab:hp_table}.

\begin{table}
 \centering
 \begin{tabular}{|c|c|c|c|}
 \hline
  Name & Range & Parameter Type & Best Model \\ 
  \hline
  Batch Normalization & {yes, no}    & Choice   & yes\\ 
  Dropout    & [0., 0.25]   & Continuous  & 0.01 \\
  LeakyReLU   & [0., 0.4]    & Continuous  & 0.15 \\
  Learning Rate  & [0.00001, 0.01]  & Continuous (log) & 0.000227 \\
  Learning Rate Decay & [0.5, 1.]    & Continuous  & 0.91 \\
  Number of Layers & [3, 12]    & Discrete   & 7 \\
  Number of Nodes  & {128, 256, 512}  & Choice   & 512 \\
  Optimizer   & {Adam, SGD, RMSProp} & Choice   & Adam \\
  \hline
 \end{tabular}
 \caption{Hyperparameter Space. The resulting best model configuration is shown in the right-most column.}
 \label{tab:hp_table}
\end{table}

This hyperparameter search took place in two stages, using ``Sherpa''~\cite{hertel2020sherpa}, a Python library for hyperparameter tuning. First, we fit a large suite (over 200) candidate DNN models using a random search algorithm. The random search has the advantage of making no assumptions about the network architecture or the task of interest. In this stage all hyperparameters, except the learning rate and learning rate decay, are modified. Excluding learning rate parameters in the first stage is strategic to ensure that any increases in performance are due to more skilled architectures.

Following the initial search, we conducted a second search on the best performing model uncovered during the first stage. This secondary investigation, which tested another fifty models, focused exclusively on the learning rate and learning rate decay settings. This procedure allowed us to train the network with the best possible learning schedule so as to maximize the networks performance while fixing the best-performing architecture uncovered in the first stage. 

In total, we tested more than 250 network architectures. We noticed a dramatic improvement in performance from the hyperparameter search quantified by the difference between the initial model's MSE and the MSE of the stage 2 model. We also observed the benefit of tuning the learning rate and the learning rate decay in stage two. The validation loss of the stage 2 model descends smoothly and consistently compared to the more archaic original model or stage 1 model. The final result of the hyper-parameter search is shown in Table~\ref{tab:hp_table}. We discuss below in the results section the extent to which hyperparameter tuning improves the benchmarks discussed above. To help quantify the improvements from the formal tuning we compare this "Best" DNN that was the result of the formal hyperparameter search against the "Manual" DNN that was designed similar to neural networks used in previous aqua-planet studies~\cite{Rasp9684}. Additionally we run tests on a baseline "Linear" model that is identical in all ways to the "Manual" DNN except for the fact that all activations are replaced with the identity function prior to training.

While it would be interesting to know whether skillful models could have been obtained with less data volume, this is impossible to precisely quantify without performing Sherpa hyperparameter tuning on a smaller dataset -- something we opted not to do due to the heavy GPU requirements of applying Sherpa. We strategically only utilized it on our richest training dataset to conserve resources. For context on the resource requirement, each candidate neural network architecture required roughly twenty-four hours to train (about one hour per epoch). Eight models could be run in parallel on a single GPU and thus, using four GPUs, we could train about thirty two models per day. In total, with four GPUs (12 GB memory each), it took eight days to train all 250 models. 

\section{Results}\label{Results}

Here we use the diagnostics outlined in Section~\ref{METRICS} to benchmark the performance of our DNNs. We quantify the overall performance of our DNNs in emulating atmospheric sub-grid heating and moistening tendencies in Sections~\ref{subsub:Spatial},~\ref{subsub:Temporal}, and analyze the emulated hydrological cycle in Section~\ref{daily_cycle}. Note that since we used the first eight years to train the network and the ninth year to optimize the hyperparameters, we benchmark the performance of our DNN on the remaining tenth year that we held out for testing. 

\subsection{Spatial Structures\label{subsub:Spatial}}

\begin{table}
 \centering
 \begin{tabular}{|c|c|c|c|c|c|c|c|}
 \hline
  Label & Training Data & Region & Variable & timestep & 25th & 50th & 75th \\ 
  \hline
  a & aqua-planet & Ocean & Heating & 15 min. & 0.05 & 0.27 & 0.55\\
  b & real-geog. (Linear) & Ocean & Heating & 15 min. & -0.30 & 0.00 & 0.21\\
  c & real-geog. (Linear) & Land & Heating & 15 min. & -0.93 & -0.06 & 0.25\\
  d & real-geog. (Manual) & Ocean & Heating & 15 min. & -0.26 & 0.00 & 0.31\\
  e & real-geog. (Manual) & Land & Heating & 15 min. & -0.93 & -0.06 & 0.35\\
  f & real-geog. (Best) & Ocean & Heating & 15 min. & 0.28 & 0.54 & 0.76\\
  g & real-geog. (Best) & Land & Heating & 15 min. & 0.41 & 0.65 & 0.82\\
  \hline
 \end{tabular}
 \caption{Statistical breakdown of skill score. We show quartiles of the skill distribution in 3D space, i.e. from a flattened vector of $R^{2}$ values that were calculated across just the time dimension separately for each longitude, latitude, and pressure level, using raw convective heating tendency data at the 15-minute sampling scale. We compare a neural network trained on aqua-planet data (a) with three different neural networks trained on more realistic SPCAM5 data. These models include a baseline "Linear" model (b-c), a manually tuned neural network (d-e), and a neural network formally tuned by Sherpa (f-g).}
 \label{tab:IQR}
\end{table}

\begin{figure}[ht!]
\centering
\includegraphics[width=\textwidth]{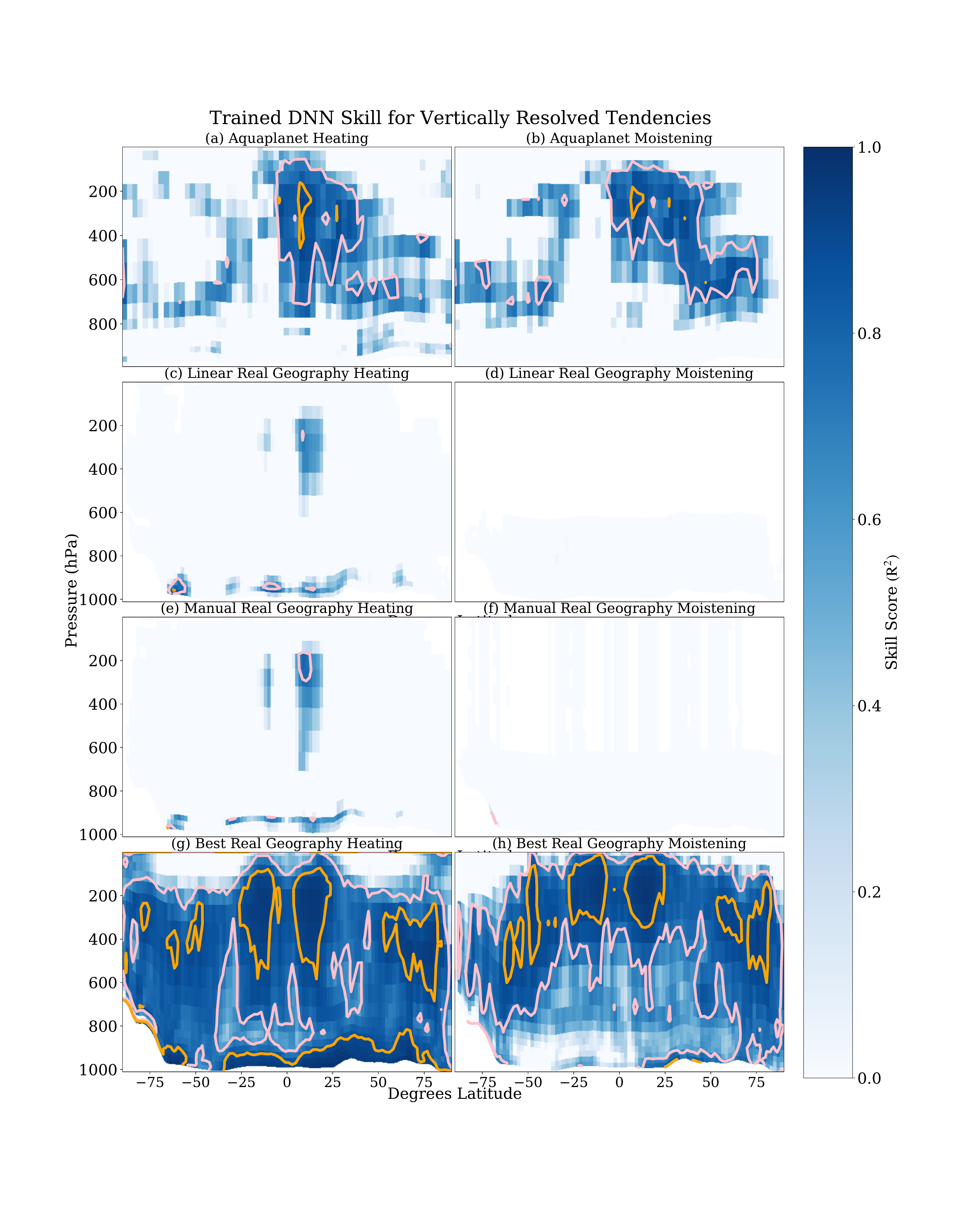}
\caption{\label{fig:SPCAM3} The ${R}^2$ coefficient of determination for zonally averaged DNN predictions. We contrast the performance of a manually tuned deep neural network emulating aqua-planet target data (a and b) with three comparable neural networks trained on full complexity real-geography data. These include our baseline linear model (c and d), a manually tuned neural network (e and f) and our semi-automated, formally tuned Sherpa neural network (g and h). Skill shown separately for heating tendency in $(K / s)$ (a, c, e, g) and moistening tendency in $(kg / kg / s)$ (b, d, f, h). Areas where $R^{2}$ is greater than 0.7 agreement between are contoured in pink and areas greater than 0.9 in orange. Convective Tendencies are zonally averaged prior to the calculation of ${R}^2$ to allow for a cross section visualization. For ease of visualization and cleaner comparison with previous work we show the plot of max(0,$R^{2}$).}
\end{figure}

Differences between the aqua-planet and full complexity real-geography test beds become clear when we analyse the performance of neural networks without any spatial averaging in statistics. While even a manually tuned feed-forward neural network can fit much of the variations in convective tendencies in an aqua-planet ($R^2 > 0.5$ at the 75th percentile-- Table~\ref{tab:IQR}a), an identical manually tuned neural network architecture performs far worse in emulating convection with land masses (Table~\ref{tab:IQR}a vs. b-e). However, the effects of hyperparameter tuning are dramatic, boosting the 50th percentile $R^{2}$ to over 0.5 and 75th percentile $R^{2}$ to over 0.75 for convective heating tendencies (Table~\ref{tab:IQR} d,e vs. f,g). The fact that one quarter of the domain of convective heating tendency is emulated with $R^{2} > 0.75$, even prior to any averaging in space or time, suggests our final DNN setup (full training data volume and hyperparameter tuning) generates a good fit. In the Supplemental Information we include corresponding statistics for convective moistening as well as for both heating and moistening tendencies on the diurnal time scale -- all of which show similar relationships between the four models. The differences in Table~\ref{tab:IQR} between land and ocean (f vs. g) indicates spatial variations in the skill, suggesting certain regions are preferentially fit by our DNNs. 

A first look at spatial structures in the skill affirms a generally close fit with familiar structures relative to aqua-planet expectations but also some interesting differences. Figure~\ref{fig:SPCAM3} presents the skill of zonally averaged DNN predictions. Here we again compare our prototype manually tuned DNN's skill on aqua-planet target data (Figure~\ref{fig:SPCAM3} a,b) representative of what was used in~\citeA{Rasp9684} against our "Linear" baseline model (Figure~\ref{fig:SPCAM3} c,d), our "Manual" DNN (Figure~\ref{fig:SPCAM3} e,f), and our optimized "Best" DNN (Figure~\ref{fig:SPCAM3} g,h) all trained in real-geography. Achieving realistic performance on zonal means is an easier objective due to the averaging between land and marine atmosphere and the smoothing of the sharpest temporal variations in convection. However, this zonal mean perspective still provides a useful composite view of the emulation of the atmosphere with land masses. Here the $R^{2}$ for zonal mean net diabatic heating and moistening is greater than 0.7 throughout the free troposphere, agnostic to latitude (Figure~\ref{fig:SPCAM3}g and h). This widespread skill in the upper troposphere is further amplified by cores of $R^{2}$ greater than 0.9 around mid-latitude storm tracks and locations of deep tropical convection above the southern and northern bounds of the ITCZ. Our "Best" DNN can skillfully emulate heating and moistening tendencies of convection across latitude and vertical level close to SPCAM5 target data (Figure~\ref{fig:SPCAM3}). It is also reassuring that the best skill ($R^{2}$ over 0.9 for zonal mean predictions) occurs in important regions of the troposphere (ITCZ, mid-latitude storm tracks) where mean diabatic heating couples to the general circulation (Figures~\ref{fig:SPCAM3}). 

When interpreting the different DNNs in Figure~\ref{fig:SPCAM3}, it is important to consider the two combined factors that can influence skill: the dataset and the quality of the DNN fit. To separate their influence, Figure~\ref{fig:SPCAM3}a,b vs. e,f shows the effect of switching training data at fixed architecture. We note a skill increase in the continental boundary layer, consistent with the existence of new deterministic signals in real-geography settings likely associated with the strong, predictable diurnal cycle over land. However, the overall atmosphere is now more complex with both land and seas regimes, as well as interactions between the two. We suspect this causes the decrease in upper-tropospheric emulation skill, particularly outside the Hadley Circulation and for the moistening tendency. Despite a superior fit in the continental boundary layer, the inability of the "Manual" neural network to capture deep convection would be a significant hurdle to online coupling. Fortunately, Figure~\ref{fig:SPCAM3}g,h shows dramatic skill improvement when migrating from a manual tuning environment to a formal hyperparameter search, underscoring the crucial role that Sherpa can play in identifying the optimal or "Best" DNN. Our preliminary analysis suggests that even under increasingly complex conditions, the "parameterizability" of convection can be cast locally in space and time for a fit by a feed-forward DNN.

These results have much in common with the findings from aquaplanet trained DNNs in~\citeA{Rasp9684} and our own aqua-planet benchmark (Figure~\ref{fig:SPCAM3}a and b) is further evidence of these similarities. However, unlike the aqua-planet, there is a new region of high skill in the real-geography emulator with $R^{2}$ greater than 0.9 in the planetary boundary layer for heating tendency emulation. This signal in convection appears deterministic enough that even our "manual" neural network can emulate it with $R^2 > 0.7$ (Figure~\ref{fig:SPCAM3}e). This looks to be a continental signal, evidenced by both higher skill in the northern hemisphere (Figure~\ref{fig:SPCAM3}h) and comparatively lower near-surface zonal mean skill at the latitudes of the Southern Ocean. Though less skillful overall, there is a similar pattern in the convective moistening tendency in the boundary layer, where the highest DNN performance at the surface ($R^{2}$ over 0.7) is confined to the continent heavy northern hemisphere (Figure~\ref{fig:SPCAM3}h).

\begin{figure}[ht!]
\centering
\includegraphics[width=\textwidth]{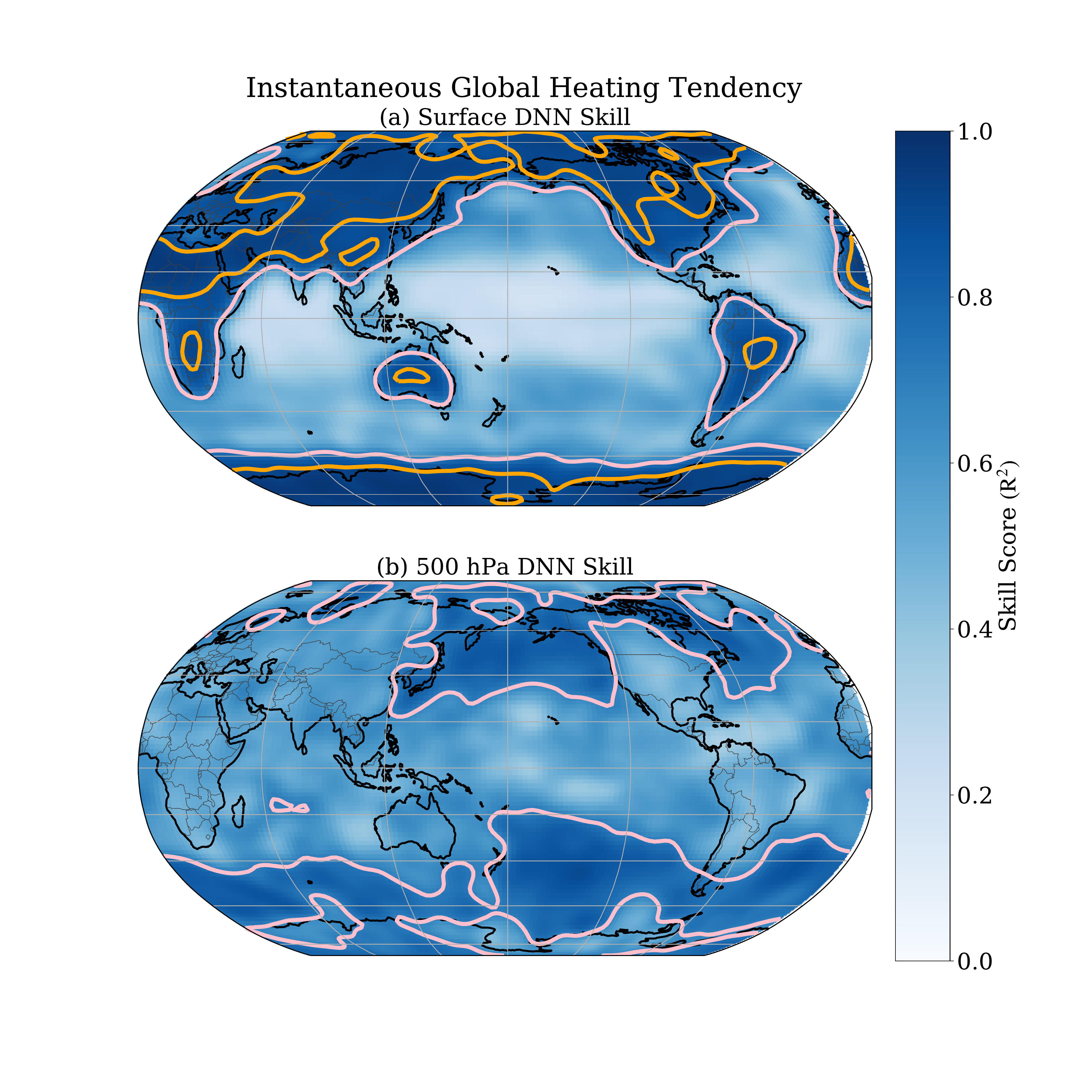}
\caption{\label{fig:R2_SFC} The neural network skill in emulating sub-grid heating at (a) the lowest model level and (b) the model level closest to 500 hPa, both at the native 15 minute timestep interval. The neural network fits locations over continents and the mid-latitudes best down at the surface, while locations of mid latitude storm tracks are best fit by our neural network in the mid-to-upper troposphere above 500 hPa. The tropics, in particular tropical locations over oceans, create the greatest challenge for the neural network emulation of sub-grid heating tendencies. Areas where the coefficient of determination $R^{2}$ is greater than 0.7 are contoured in pink and areas greater than 0.9 in orange. To facilitate reading, the map was smoothed using a 2D Gaussian averaging kernel with a standard deviation of 2 grid cells in both latitude and longitude (y and x). Each Gaussian filter was additionally truncated at 4 standard deviations. For ease of visualization and cleaner comparison with previous work we show the plot of max(0,$R^{2}$).}
\end{figure}

We confirm the existence of some distinct land-sea spatial structures in emulation skill by examining maps of predictions prior to any spatial averaging. At the lowest model level, our "Best" DNN predictions achieve $R^{2}$ greater than 0.7 (greater than 0.9 in continental interiors; Figure~\ref{fig:R2_SFC}a). However, this spatial pattern is inverted when examining skill on a model layer in the midtropopshere, near 500 hPa. At this altitude, our "Best" DNN now makes the most accurate predictions over the extratropical marine atmosphere but struggles over continents and deep convecting regions of the tropics (Figure~\ref{fig:R2_SFC}b). We speculate that a strong, deterministically predictable component of diurnal variability in surface heating and moistening associated with large surface flux diurnal variations over land could allow the low-level heating skill to be enhanced there. Meanwhile, in the upper troposphere we see the expectd to see skill deficits in regions of tropical and continental convection (on this 15-min timescale). Diurnal signals will be examined in greater detail in Section~\ref{sec:diurnal}.

We now focus on the spatial structures where even the formally optimized "Best" neural network still struggles. For this we return to assessing zonal mean predictions, since these are less exposed to details of stochasticity but are particularly important to emulate accurately when using neural networkss prognostically. The greatest emulation challenge for our "Best" DNN is fitting mean temporal variance throughout the lower troposphere (excluding the continental boundary layer) where $R^{2}$ falls below 0.3 (Figure~\ref{fig:15min_vs_daily_zonalmean}a). This is especially challenging in the case of convective moistening (Figure~\ref{fig:15min_vs_daily_zonalmean}b). Our results here are consistent with previous aqua-planet simulations~\cite{Rasp9684, Gentine_all}, and the study of~\citeA{hanetal2020}. Boundary layer moistening is an especially challenging target for machine learning emulation, particularly when focusing on the 15-minute sampling interval. Further evidence of this challenge can be observed by the temporal standard deviations of the heating and moistening tendencies, where much of the spatial field is emulated well, but our neural network nevertheless under-predicts values of moistening tendency in the lower troposphere (Figure~\ref{fig:time_standard_deviation}).

We conclude with an animation demonstrating unfiltered, non-composited views of the convective tendency emulation of over a two week period in July, the link to which can be found in the Supplemental Materials (Movie SI 1). The animation shows the evolution of total diabatic heating and moistening on a model level near 600 hPa (the lower-to-mid troposphere). In the three lower panels, the diurnal cycle of peak nocturnal radiative cooling can be seen propagating from east to west tracking the earth's rotation. It is punctuated by local features of positive diabatic heating from latent heating within slow moving weather systems, as well as the stationary lagged diurnal convective response to the passage of the sun over Central Eurasia and America. No geographic distortions of synoptic disturbances are detectable -- even heating tendencies from tropical convective clusters and the Atlantic Convergence Zone and Pacific ITCZ are all closely emulated from this perspective. On the three upper panels the associated moisture perspective provides an especially clear view of the lack of lower amplitude motion captured in emulated convection. The main distortion compared to truth is the lack of stochasticity, which manifests as geographic static in the benchmark test data (center panels) but is absent in the DNN emulation (right panels). 

\par


\subsection{Temporal Variability\label{subsub:Temporal}}

\begin{figure}[ht!]
\centering
\includegraphics[width=\textwidth]{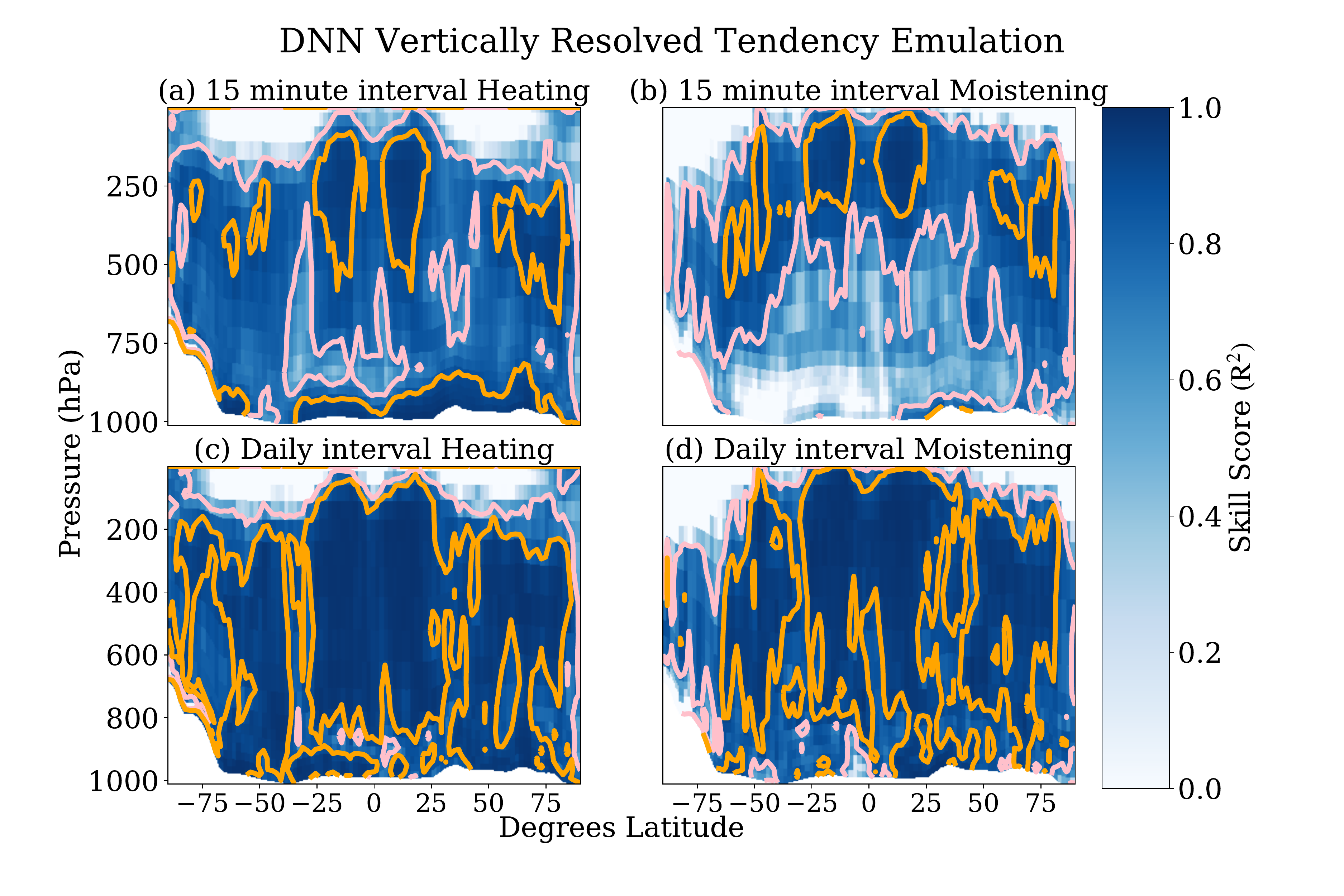}
\caption{\label{fig:15min_vs_daily_zonalmean} Neural network performance at time step interval (a and b -- also seen in Figure~\ref{fig:SPCAM3} g and h) is contrasted with performance at the diurnal scale (c and d). Representation of heating tendency in $(K / s)$ (a and c) and moistening tendency in $(kg / kg / s)$ (b and d) are both examined. Zonal averages are again taken upstream of $R^{2}$ calculation. In both vertically resolved heating and moistening, there is an across the board gain in skill at longer timescales. Areas where $R^{2}$ is greater than 0.7 are contoured in pink and areas greater than 0.9 in orange. For ease of visualization and cleaner comparison with previous work we show the plot of max(0,$R^{2}$).}
\end{figure}

Why do we see such considerable variations in the skill of our DNN as a function of geographic location and altitude? One hypothesis is that the DNN fits ``mode-specific'' fluctuations. A first test of this is re-examining spatial skill structures after averaging predictions from their native timescale of 15 minutes to the daily mean timescale instead. Figure~\ref{fig:15min_vs_daily_zonalmean} shows the corresponding skill for daily-mean, zonal-mean predictions. From this view, the vast majority of the atmosphere can be emulated in terms of both heating and moistening tendencies with $R^{2}$ greater than 0.9. Meanwhile compared to the faster timescale, the skill deficit in the lower troposphere for convective moistening tendency is not nearly as dramatic. The fact that structures in spatial skill appear sensitive to temporal averaging is consistent with the hypothesis that the DNN performance might be "mode-specific". 

\begin{figure}[ht!]
\centering
\includegraphics[width=\textwidth]{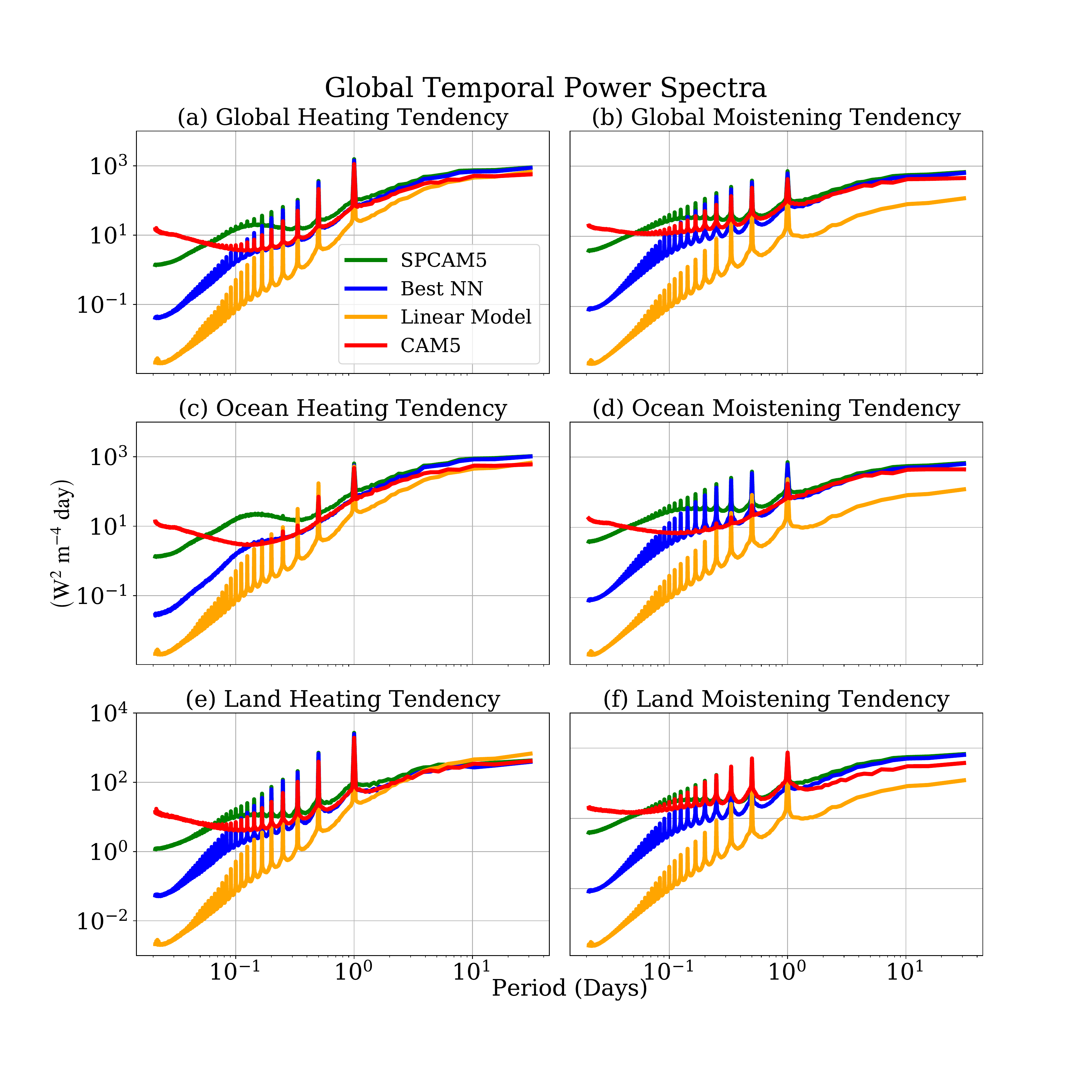}
\caption{\label{fig:spectra} The temporal power spectrum for vertically resolved heating tendency in $(W^{2} / m^{4} day)$ (a) and vertically resolved moistening tendency in $(W^{2} / m^{4} day)$ (b) are calculated at each latitude, longitude, and elevation across the globe. These spectra are then averaged together to see how much variance the linear baseline model captures globally compared to our formally tuned Sherpa neural network. Results from SPCAM5 test data and CAM5 data are also plotted for perspective. Further tests are done exclusively over marine locations (c and d) and over continental ones (e and f). The peaks correspond to the solar radiation driving the diurnal cycle, though this is stronger on land (e and f) than in marine locations (c and d). Multi-taper spectra were also calculated for both tendencies but showed no qualitative difference with the results above calculated through the numpy fft package.}
\end{figure}

To better answer understand whether our neural network only fits a dominant mode or two of convection at the expense of lower amplitude variations, we now turn to spectral analysis (Figure~\ref{fig:spectra}) on the SPCAM5 target data and DNN predictions. Switching to frequency space is a clean way to determine if specific modes of variation such as the diurnal heating cycle and synoptic storm propagation are driving preferential modes of DNN fit. The PSD is calculated separately at each unique latitude, longitude, and vertical level from the CAM5 data, SPCAM5 data, and the corresponding DNN output for both our Best neural network and "Linear" baseline model, and weighted by layer mass. These location-specific PSD are then averaged together horizontally and vertically to arrive at a globally representative power spectra.

In contrast to our hypothesis, the spectral analysis does not reveal any major mode-specificity in the DNN skill on the hourly-to-weekly timescale. All of the most important spectral features in the target data exhibit comparable power in the DNN predictions, including the main signals from disturbances slower than one day, but also the discrete variance from diurnal, semi-diurnal, and other harmonics of the daily cycle of convection. While there is an expected under prediction of total variance for sub-diurnal modes, the DNN skill is not obviously preferential to any modes at the diurnal timescale or longer. 

\begin{figure}[ht!]
\centering
\includegraphics[width=\textwidth]{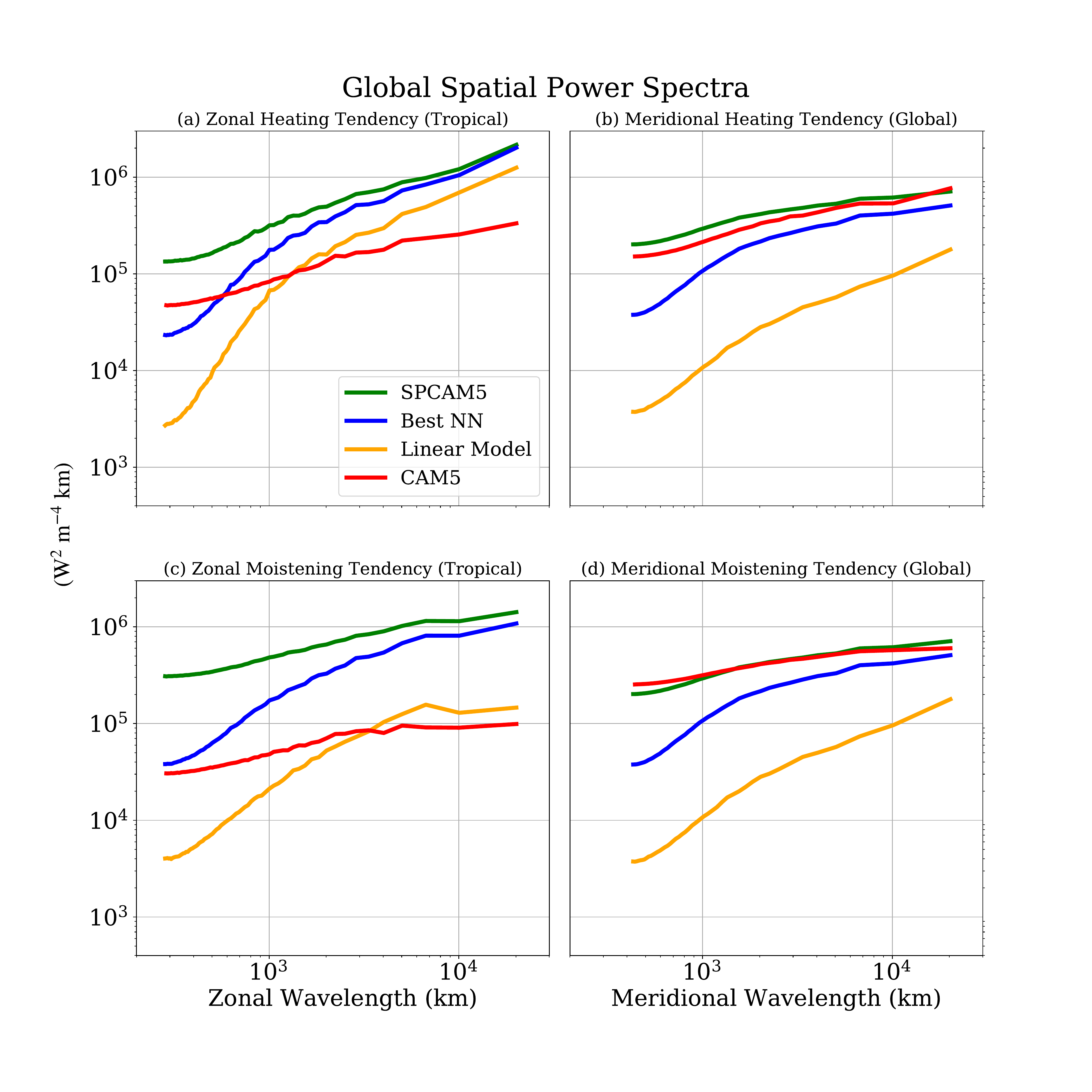}
\caption{\label{fig:space_spectra} The spatial power spectrum for vertically resolved heating tendency in $(W^{2} / m^{4} km)$ (a and b) and vertically resolved moistening tendency in $(W^{2} / m^{4} km)$ (c and d) are calculated at each vertical level and time step across the simulation data. These spectra are then averaged together to see how much variance the linear baseline model captures globally compared to our formally tuned Sherpa neural network. Results from SPCAM5 test data and CAM5 data are also plotted for perspective. We take a 1D fft in both the x (zonal) (a and c) and y (meridional) (b and d) directions. However, we restrict our zonal cross-section to just a tropical belt (20N-20S) so we can assume a cartesian plane and neglect variable grid spacing. These results tie in with Figure~\ref{fig:spectra} in that capturing the variations in convective tendencies at small scales proves more difficult for our neural networks than at large scales.}
\end{figure}

Our DNN emulation performance can be further analyzed by taking the PSD from a spatial, rather than temporal, domain. Here the DNN also shows skill at capturing variations in convection at large scales but does not emulate all the details at the small spatial scales (Figure~\ref{fig:space_spectra}). We note that this is very much in line with the findings of~\citeA{Yuval_Ogorman} in which their Random Forests achieved poorer fits at smaller spatial scales compared to wider ones when performance was tested on different course-graining length scales.

While even our simple baseline "Linear" model (orange line in Figure~\ref{fig:space_spectra}) can capture the variance in convective tendencies on a global scale, we see evidence of the benefits of automated, formal hyper-parameter at the model grid cell scale. While our "Best" neural network still underestimates smaller signals in convection, it is much closer to the SPCAM5 test data with respect to both convective heating and moistening. 

We can quantify these differing degrees of skill captured in the spatial spectra by calculating the total log spectral difference (LSD). Quantitatively, the LSD between the SPCAM5 target data and the Sherpa "Best" neural network predictions is 1.19 for mass-weighted, averaged tropical zonal heating tendency spectra and 1.55 for mass-weighted, averaged tropical zonal moistening tendency spectra from Equation~\ref{eq:LSD}. This is a far smaller deviation than when the baseline "Linear" model is compared to the target SPCAM5 data and the difference between the tropical averaged, mass weighted zonal spectra are 2.71 and 3.61 for heating and moistening respectively. Reassuringly, our "Best" neural network also has a lower LSD than the difference between CAM5 and SPCAM5 data (1.20 and 2.30 for heating and moistening zonal spectra). From the temporal domain, the "Best" neural network has a far smaller LSD for both the heating and moistening spectra than the "Linear" baseline. However, the LSD between the CAM5 and SPCAM5 is actually smallest (in the temporal domain), though from the figure it is clear that this is due to behavior at the shortest time scales that produce exponentially less variance, but which are up-weighted by this metric (not shown). 

Even with the sophisticated hyperpameter tuning, the very weak signals of fast variability on timescales less than 2-6 hours is where our DNN is still challenged most (Figure~\ref{fig:spectra}). Evidently there is something native to high (spatial or temporal) frequency heating and moistening convective tendencies that challenges our DNN. This suggests an alternate hypotheses that geographic structures in our DNNs skill might be an artifact of variance sorting (i.e. the fact that the fastest signals are also the weakest potentially downweights their contribution to the loss function) or of stochasticity (since fast variations can often be stochastic in origin). 

\begin{figure}[ht!]
\centering
\includegraphics[width=\textwidth]{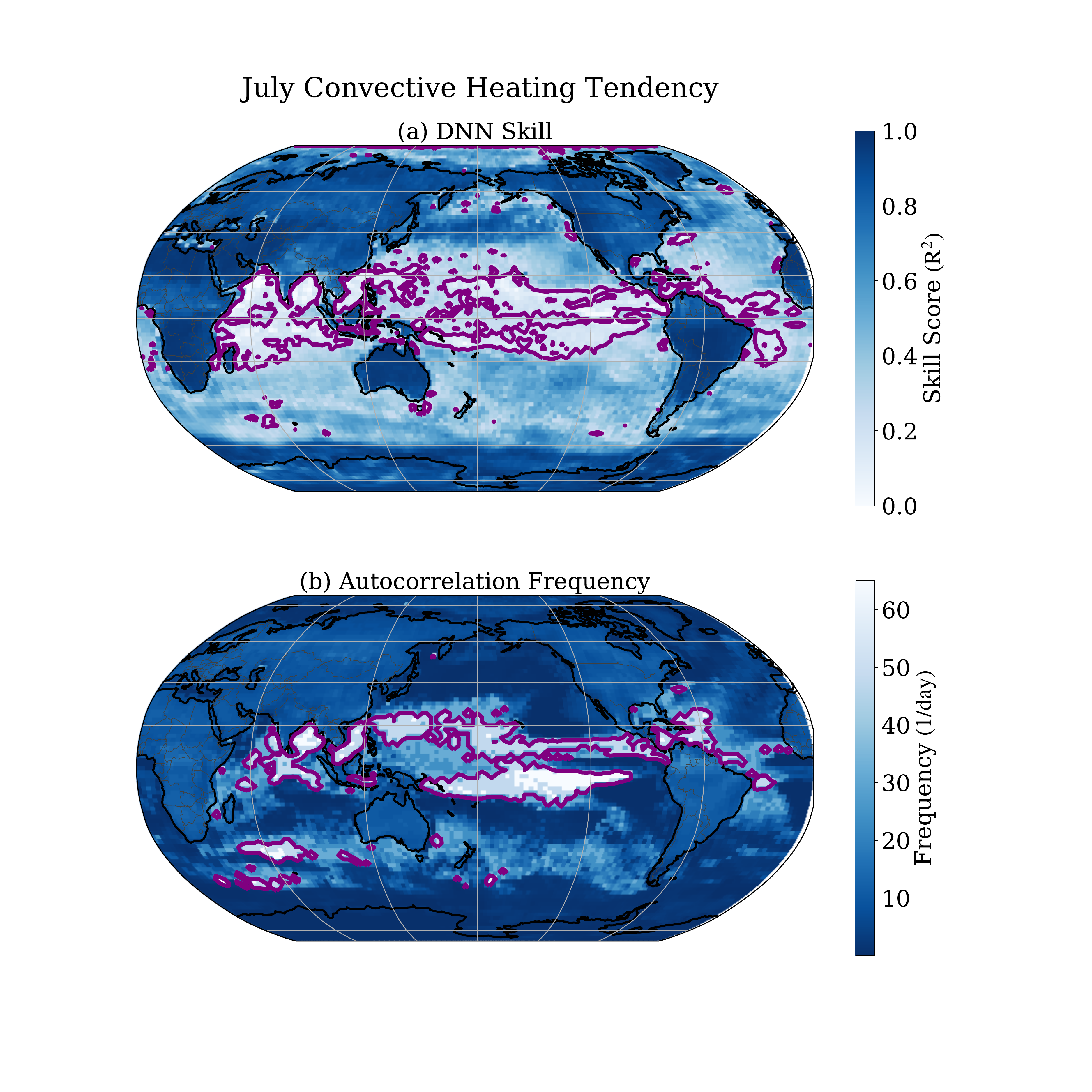}
\caption{\label{fig:Autocorr} A comparison between the neural network ${R}^2$ skill in emulating the vertically resolved heating tendency in $(K / s)$ (a) and the autocorrelation frequency of the SPCAM5 heating tendencies (b). Both cross sections are taken at the lowest pressure level in the model. Qualitatively the patterns closely match. The areas of lowest skill score (bottom tenth percentile) and highest autocorrelation frequency (90th percentile) are both contoured in purple. For ease of visualization and cleaner comparison with previous work we show the plot of max(0,$R^{2}$) in panel a.}
\end{figure}

\begin{figure}[ht!]
\centering
\includegraphics[width=\textwidth]{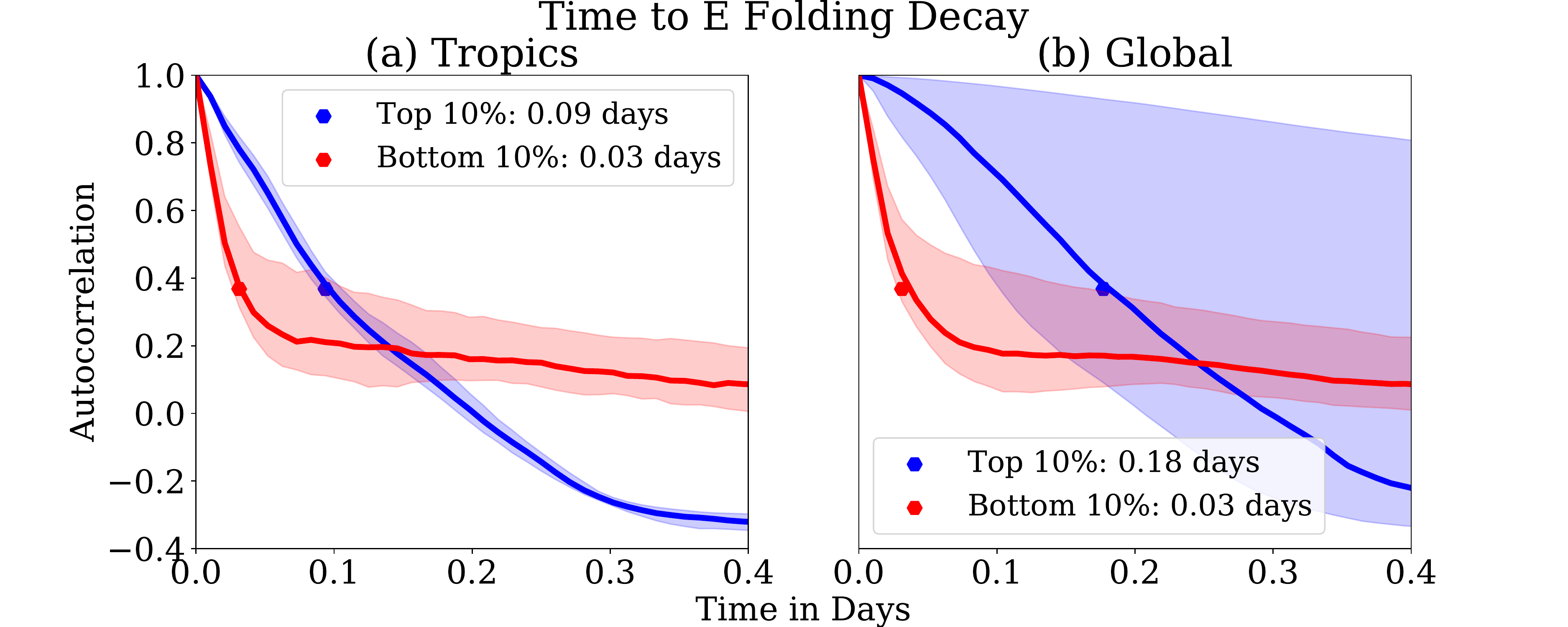}
\caption{\label{fig:Autocorr_Regions} The solid lines represent the the median autocorrelation as a function of time at every surface location where the ${R}^2$ skill score of heating tendency in $(K / s)$ is in the top 10 percent (blue) and the bottom 10 percent (red). We restrict our comparison to surface locations in the tropics ($15^{\circ}$S to $15^{\circ}$N) (a) and then examine the entire surface of the earth (b). The corresponding inter-quartile regions are shaded in as a marker for statistical significance. The dots show the time to e-folding decay. The test data spans the month of July.}
\end{figure}

To further test these hypotheses, we construct a proxy of stochasticity, and compare its geographic structure to the geographic structure of DNN ${R}^2$ skill. The proxy is the e-folding time at which signals of atmospheric variance decouple into noise based on the autocorrelation, i.e. the decorrelation timescale. Quantitatively, the spatial structures of ${R}^2$ heating skill and the de-correlation timescale are similar with pattern correlation coefficient of 0.50. The fact that the regions of least DNN skill are also the regions of fastest signal decorrelation (purple contours in Figure~\ref{fig:Autocorr}) supports the view that imperfect emulation of fast, stochastic signals is mainly responsible for sculpting the spatial structures in the DNN's skill score. 

To illustrate this further within the challenging tropical regime, Figure~\ref{fig:Autocorr_Regions} examines temporal autocorrelations in different DNN skill regimes as follows: First, all tropical grid cells having poorest skill (the bottom 10 percent) are identified, and the temporal autocorrelation of the benchmark time series data is calculated from time lags 0 to +0.4 days. This is done separately for each tropical grid cell and then composited into a single autocorrelation plot (red line). Repeating the procedure for those horizontal tropical grid cells where the DNN fit has highest skill (top 10 percent, the blue line) reveals the characteristic difference in de-correlation timescale in the high-skill vs. low-skill spatial regions. Repeating this procedure globally (Figure~\ref{fig:Autocorr_Regions}b) confirms the same timescale-selectively of skill exists across multiple geographic regimes, even though this was not obvious in Figure~\ref{fig:Autocorr}. The relationship is robust and clear -- locations where signals tend to decorrelate faster are locations where DNN skill is lower. This is consistent when examining both the planetary boundary layer of the oceans and the continents each in isolation as well as globally (not shown).

\subsection{Hyperparameter Optimization vs. Physical Constraints for Emulating the Diurnal Cycle}\label{daily_cycle}\label{sec:diurnal}

\begin{figure}[ht!]
\centering
\includegraphics[width=\textwidth]{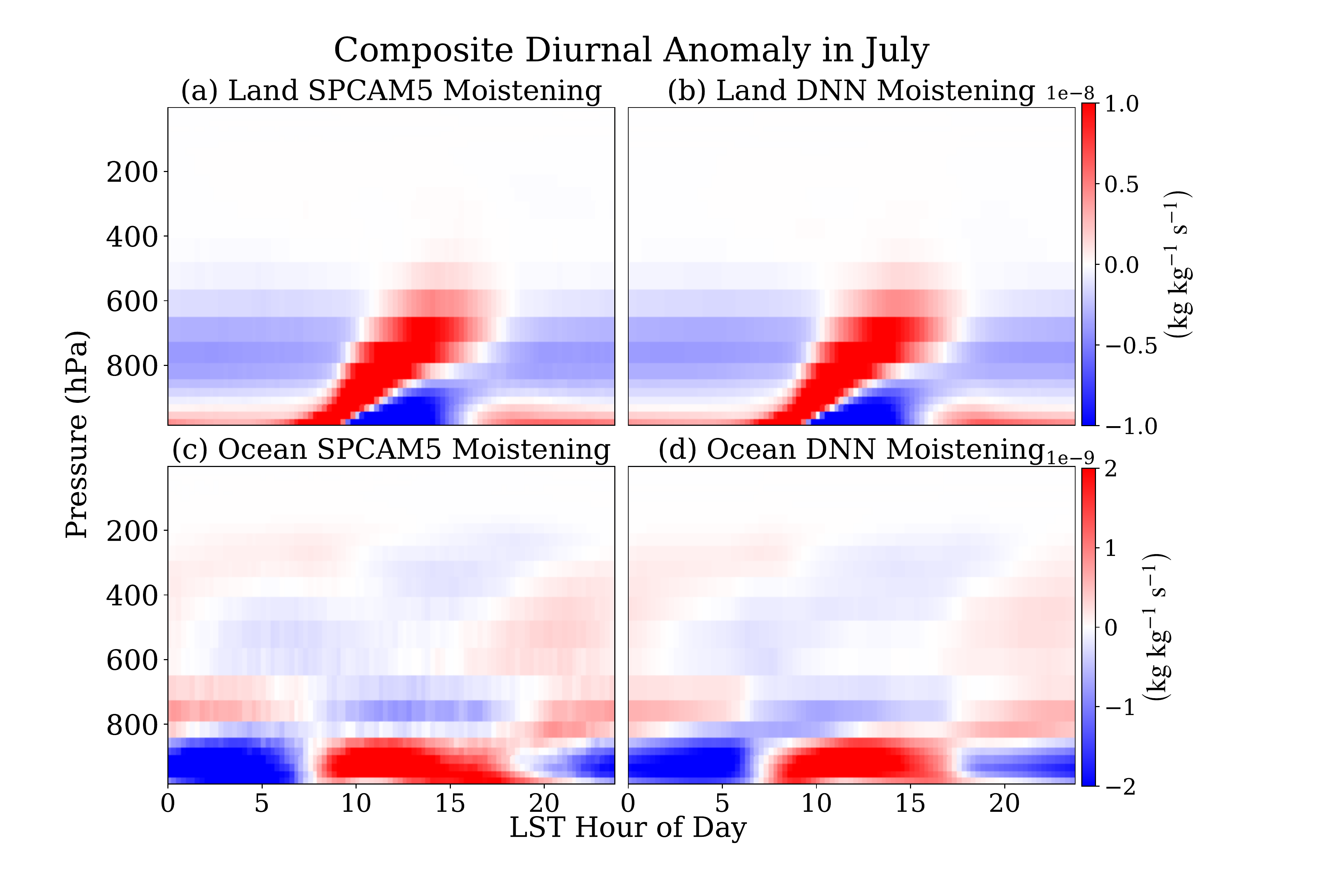}
\caption{A comparison between the moistening tendency of SPCAM5 target data (a and c) and DNN predictions (b and d) in $(kg / kg / s)$ over continental (a and b) and marine (c and d) locations respectively. The composite is taken over the month of July and we choose to show the anomaly of the diurnal cycle in which the mean is subtracted out.}
\label{fig:diurnal_height_time}
\end{figure}

\begin{figure}[ht!]
\centering
\includegraphics[width=\textwidth]{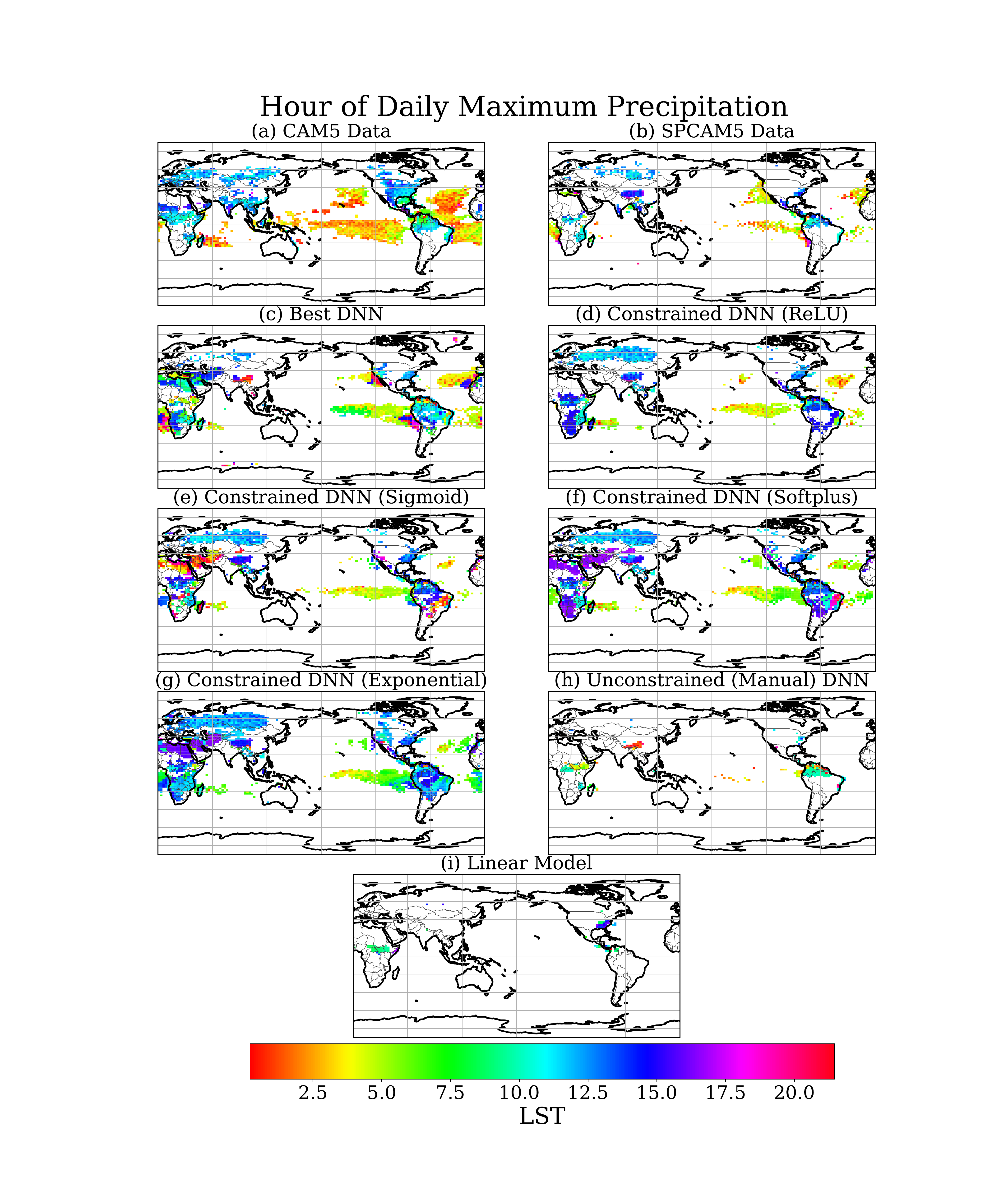}
\caption{\label{fig:diurnal_phase_map} A comparison between CAM5 data (a), SPCAM5 test data (b), and our overall best neural network with automated hyperparameter tuning (c), neural networks with different positive constraints on the precipitation output (d, e, f, g), an archaic version of our DNN without automated hyperparameter tuning or physical constraints (Manual) (h), and our linear baseline model (i). The figures show the hour of maximum precipitation in $(mm / day)$ during the boreal summer (months of June, July, and August). The time of maximum precipitation is colored in only over areas with a significant diurnal amplitude in precipitation rate as defined in Equation~\ref{Precip_Sig}.}
\end{figure}

So far we have shown comprehensive skill for our optimized DNN except for some fast-varying convective signals that de-correlate quickly. On the one hand it is not obviously a problem to have skill deficits for the stochastic component of superparameterized convection, since it does not appear to be critical to most emergent behaviors of SPCAM~\cite{Randall_2019_Ensemble} -- though some have suggested that a close representation of the stochastic component is still necessary to properly model large scale weather phenomena~\cite{Neelin_stochasticity}. On the other hand, some fast-varying signals are also critical to regional climate simulation and should be deterministically predictable. The diurnal cycle should provide us a perfect test-bed. To what extent does our DNN emulate the details of the diurnal cycle of convection? 

A first look at the composite height vs. time-of-day structure of convective moistening (Figure~\ref{fig:diurnal_height_time}) is reassuring -- the DNN captures the coherent temporal transition of shallow to deep convection in the afternoon over land (moistening above drying growing after sunrise into the mid-troposphere; Figure~\ref{fig:diurnal_height_time}a vs. b). Our DNN also shows a good fit over the ocean, where it captures the opposite phase of peak moistening-above-drying which happens during the night between 8pm and 6a (Figure~\ref{fig:diurnal_height_time}c vs. d). Thus our first impression is that the DNN correctly recreates the land-sea contrast of diurnal convection present in the SPCAM5 target data (Figure~\ref{fig:diurnal_height_time}). 

We now hone in on the full geographic structure, focusing on the diurnal cycle of \textit{precipitation}, which reveals some interesting surprises. The benchmark SPCAM5 target data (Figure~\ref{fig:diurnal_phase_map}b) resembles observations. Over land regions our test data shows a strong, predictable diurnal precipitation cycle over tropical rainforests and continents in the northern hemisphere (experiencing boreal summer), with lagged afternoon maximum precipitation (Local Solar Time between 13:00 and 18:00). In contrast, a weak diurnal cycle of precipitation occurs over the oceans that peaks at the end of night into early morning, and is especially detectable in subtropical stratocumulus regions. We observe the familiar benefits of superparameterization relative to conventional parameterization (Figure~\ref{fig:diurnal_phase_map}b vs. a) including a reduced detectability of the diurnal mode except where it is supposed to be strong such as over tropical rainforests or where it is especially consistent such as over marine subtropical drizzle regimes. Here we see an interesting result in the optimized DNN's precipitation predictions (Figure~\ref{fig:diurnal_phase_map}c): Although the stratocumulus marine drizzle cycle appears to be well emulated, consistent with the diurnal moistening composites seen in Figure~\ref{fig:diurnal_height_time}d, over land there is incorrect timing and spatial extent of maximum precipitation (Figure~\ref{fig:diurnal_phase_map}c). This is paradoxically at odds with Figure~\ref{fig:diurnal_height_time}a and b which indicated excellent emulation of the diurnal cycle of convective moistening over land regions. DNN detection of a cycle of precipitation over desert regions is physically unrealistic and the timing of the onset of deep convection and heavy precipitation on land is several hours premature, much like CAM5 data (Figure~\ref{fig:diurnal_phase_map}c, b, and a). 

Why is precipitation emulated less skillfully? Our working hypothesis is that in hindsight our DNN architecture did not respect an important physical constraint that distinguishes this variable. Unlike moistening and heating tendencies, precipitation should be positive definite. To test this hypothesis we introduce four additional neural networks in Figure~\ref{fig:diurnal_phase_map} (d,e,f,g). Each new neural network has a different positive constraint (nonlinear activation function) on the last precipitation node to ensure rainfall predictions remain positive definite in line with physical reality. Additionally, in these new constrained DNNs we alter the training data used. We require less data overall (just three months) but empirically find that we should no longer selectively sample as we did previously (e.g. take a day every 10 days). We note that restricting the training dataset to less than a full year does not seem to cause problems with out of sample test data; it emulates the diurnal cycle of precipitation just was well over boreal winter even when the training data is restricted to boreal summer (not shown). We compare our previous DNNs (Best, Manual, Linear baseline model) that ignored positive definite nature of precipitation but included differing hyperparameter tunings (Figure~\ref{fig:diurnal_phase_map}c,h,i). against these new positively constrained, but not formally tuned networks. 

Our results show the different positive constraints induce improvements over different regions of the globe. However, there are substantial variations between different constraint choices with little systematic effects other a realistic enforcement of the timing of the onset of maximum precipitation over land and (except ReLU) a tendency towards poor emulation in dry regions of Africa and the Middle East where these three constrained (by Softmax, Exponential, and Sigmoid functions) neural networks invent a diurnal cycle of precipitation. Unsurprisingly, the DNNs with neither positive constraints nor formal hyperparameter tuning (Figure~\ref{fig:diurnal_phase_map}h and i) perform the worst. The timing of maximum daily precipitation is premature over land: noon-centric rather than peaking in the mid-afternoon (Figure~\ref{fig:diurnal_phase_map}h and i). Also, these neural networks fail to detect a diurnal cycle of precipitation over much of the globe (Figure~\ref{fig:diurnal_phase_map}b vs. h,i). Adding the positive-definite constraint alone produces dramatic improvements over land (Figure~\ref{fig:diurnal_phase_map}d-g vs. h,i)-- but there is the aforementioned variation in the magnitude of improvement between choices of activation function as a constraint. The DNN with a ReLU activation on precipitation seems to emulate the diurnal cycle of precipitation the best (Figure~\ref{fig:diurnal_phase_map}d). The hour of maximum precipitation is correct and the neural network emulates a diurnal cycle only where it should be strong, over tropical rain forests, the Southeast United States, and mid-continental summertime Eurasia (Figure~\ref{fig:diurnal_phase_map}d vs. e,f,g). Unlike the other three constrained neural networks (Figure~\ref{fig:diurnal_phase_map}e,f,g) and our Sherpa neural network (Figure~\ref{fig:diurnal_phase_map}d), it does not fabricate diurnal precipitation over the deserts of north Africa and the Middle East -- but nevertheless its emulation of precipitation over continental Africa is still imperfect. But taken on balance, a constrained (by ReLU) neural network appears to solve most of the problems that our original DNNs suffered over land. However the positive constraint alone is unable to emulate the more subtle marine stratocumulus diurnal cycle well in both the Atlantic and Pacific oceans where the Sherpa DNN emulates the correct time and spatial coverage of this lower amplitude cycle of precipitation (Figure~\ref{fig:diurnal_phase_map}c vs. d).

We have highlighted the power of automated hyperparameter tuning on convective tendencies in Figure~\ref{fig:SPCAM3}, but discovered that even our "Best" Sherpa DNN did not take into account physical laws governing its simultaneous prediction of precipitation, instead corrupting it. Difficulty emulating details of precipitation cycles are certainly not unique to this study but do point to larger growing pains in the machine learning and climate science communities. Similar problems with capturing the physics behind precipitation through neural networks have been discussed in~\citeA{Nature_ReLU, Brenowitz_Bretherton} where neural networks created non-trivial negative precipitation as well. As in~\citeA{ReLU_fix}, we show that augmenting our DNN with a positive constraint could reduce the errors in land precipitation emulation (Figure~\ref{fig:diurnal_phase_map}b vs. d). Without formal hyperparameter tuning, these constrained DNNs emulated the land-sea contrast in the timing of peak precipitation: nocturnal over oceans, late-afternoon over the hottest and moistest continental regions. In hindsight it would be logical to complement the benefits of hyperparameter tuning with such constraints -- an important topic for future work. It is also possible that skill in the precipitation field would benefit from enforcing a consistency between it and the column moistening that is better emulated, as in~\citeA{beucler2019enforcing} or~\citeA{Watt_Meyer21}. 

\begin{figure}[ht!]
\centering
\includegraphics[width=\textwidth]{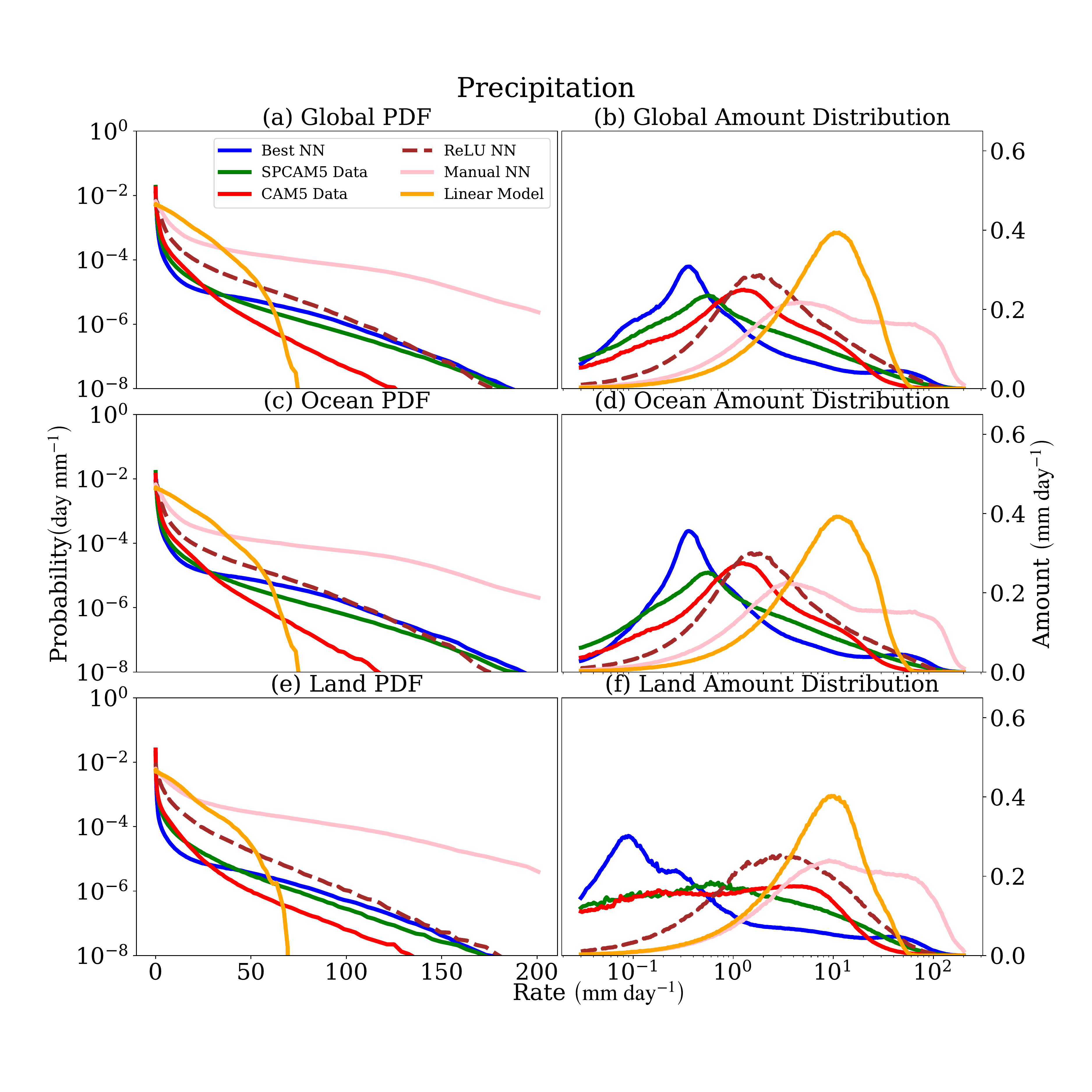}
\caption{\label{fig:Precip_Dist} The Probability Density Function across the range of simulated precipitation rates (a) and the corresponding amount distribution (b) of precipitation in which the probability density function is multiplied by the bin-averaged values of precipitation. We design the histograms based on the methods outlined in~\citeA{Watterson_Dix}, which have been widely adopted in literature including in formative works such as~\citeA{Pendergrass_Hartmann}. We implement logarithmically distributed rain-rate bins. In our case each bin width grows by 3 percent to ensure the entirety of the precipitation PDF is reflected. For more detail, we include an archaic version of our neural network without an automated hyperparameter tuning or physical constraints (Manual), our best constrained neural network (dashed line), and our overall best (Sherpa) DNN discussed previously in the methods section. Comparisons are also made exclusively over marine areas (c and e) and continental ones (d and f).}
\end{figure}

To further assess these trade-offs we now look beyond just the diurnal cycle to examine the full PDF of precipitation across our sensitivity tests. The formally tuned "Best" DNN does outperform all other neural networks in capturing the global precipitation amount distribution (Figure~\ref{fig:Precip_Dist} blue vs. green). Consistent with the diurnal cycle analysis this "Best" DNN performs especially well over the ocean (Figure~\ref{fig:Precip_Dist}d blue vs. green). However, issues over land are even more striking from the viewpoint of the full PDF where the DNNs have radically different values for the amount mode -- i.e. rainfall rate delivering maximal precipitation. Whereas the diurnal cycle analysis had suggested a positive definite constraint alone brought continental precipitation into focus, we can see from the amount distribution that beyond diurnal timing its statistics are incorrect (dashed line vs. solid green). In fact, our constrained neural network has a more accurate PDF over ocean despite its established struggle fitting the nocturnal cycle of precipitation over the marine stratocumulous regions of the globe (Figure~\ref{fig:Precip_Dist}d and Figure~\ref{fig:diurnal_phase_map}). Meanwhile the formally tuned DNN has a pronounced problem of producing too much drizzle over land which is also a problem seen in precipitation from standard parameterization. 

Taken as a whole our precipitation results suggest that this is an area where further refinement of even our "Best" DNN is needed. Over the oceans, the DNN captures much of the PDF of precipitation (Figure~\ref{fig:Precip_Dist}), including moderate to heavy regimes at the tail that challenge many climate models, as well as the diurnal cycle of precipitation over the oceans (Figure~\ref{fig:diurnal_phase_map}b vs. c). But there is substantial corruption of the emulated signal over continental locations, particular with regards to the timing of onset of heaviest precipitation and the intensity of rain delivering most surface accumulation. For an even more information-rich view, we have attached as Supplemental Information an animation showing two weeks of July precipitation from CAM5 data, target SPCAM5 data, and DNN emulation (Movie SI 2) as well as a version of Figure~\ref{fig:Precip_Dist} with all four constrained DNNs (Figure SI 1). 

Comparing the fit sensitivity of adding a constraint vs. leveraging hyperparameter optimization methods~\cite{tomsetal2020}, both methods provide unique, disparate performance enhancements with the Constrained DNN performing better over land and the "Best" Sherpa DNN doing better overall (Figures~\ref{fig:diurnal_phase_map} and~\ref{fig:Precip_Dist}). But synthesizing both tools may ultimately be necessary since neither a physically constrained neural network architecture nor automated hyperparameter tuned network on its own could capture the full complexity and timing of the diurnal cycle of precipitation over both land and ocean. We recommend an integration of both these tools for future attempts of this work. Meanwhile it is worth recalling that these corruptions are less obvious in the diurnal cycle of heating and moistening which is better emulated, perhaps because it dominates the loss function, or perhaps because -- unlike for precipitation -- there is no internal inconsistency with the values of these variables and the assumed DNN architecture. But since precipitation is a critical input to land surface models, resolving the issues revealed in this section will be an important next step towards realizing successful prognostic behavior. Other issues at the frontier of coupling emulators to land surface models are discussed next.

\subsection{Towards Interactive Land Coupling}\label{Remarks}

\begin{figure}[ht!]
\centering
\includegraphics[width=\textwidth]{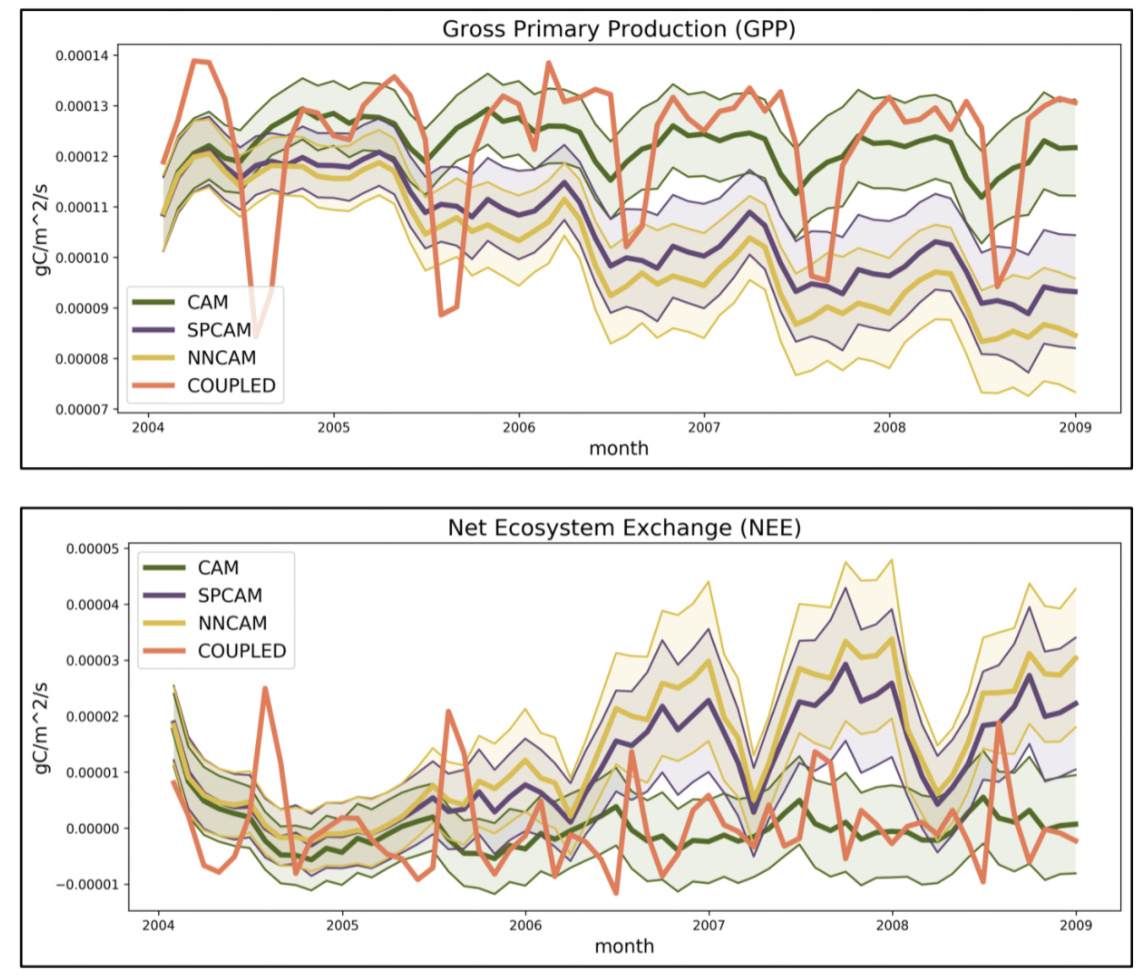}
\caption{\label{fig:Galen_Y} The Gross Primary Production (GPP) and Net Ecosystem Exchange (NEE) monthly based on CAM data (also in aqua-planet mode) are contrasted against SPCAM (aquaplanet) and a neural network (aquaplanet trained), the results of which are derived from one way land coupling. The solid lines correspond to mean values while the shading encompasses the extent of the monthly mean standard error at each time step.}
\end{figure}

Taken together, most of the above results look promising enough that it is natural to wonder if prognostic tests using an emulator like this might produce stable simulations as was shown for an aqua-planet by~\citeA{Rasp9684}, but in a real-geography setting. This would be exciting to test but our view is that as yet it is premature to try. For instance, beyond its corruptions of continental precipitation, the DNN we have described does not predict everything that would be needed to drive an interactive land surface model in practice. It is even unknown whether the imperfections in the near-surface state of the DNN's predictions would even be compatible with land surface modeling. 

As a first credibility test on the latter front we thus report some results from ``offline'' standalone land surface model simulations driven with actual (vs. emulated) surface state data. These simulations predate the real-geography fits here but use the downwelling surface solar radiation, precipitation, surface pressure, near-surface humidity and temperature, as well as wind speed from a previous neural network powered aqua-planet GCM to drive several land model integrations. These simulations are easier to perform than fully interactive land-atmosphere coupled simulations and provide a quick test of the null hypothesis that corruptions of the surface state by the DNN might be incompatible with land modeling in general. The idealized offline land model test-bed assumes Amazon-like properties, and 5-year simulations are repeated for 112 separate grid cells driven by atmospheric inputs spanning the tropical band~\cite{Yacalis2018ArtificialNN}. 

The neural network used here is philosophically similar in design to the architecture to the we use in Figure~\ref{fig:SPCAM3} (a and b). Among these similarities are the training data, which is also of T42 spectral truncation and a native 30 minute model time step. However, this neural network is trained on a full year of simulation data, rather than three months. It is also a larger neural network with 8 hidden layers of 512 nodes each and an input vector of 124 (the dynamic tendencies of temperature and humidity over the entire column are included for an extra length of 60 in the input). Likewise, the output vector includes the longwave and shortwave tendencies over the column for a total vector size of 120. The complete details can be found in~\citeA{Yacalis2018ArtificialNN}. 

The results in Figure~\ref{fig:Galen_Y} reaffirm the potential for prognostic tests. Figure~\ref{fig:Galen_Y} shows the carbon cycle flux responses from the resulting ensemble of Community Land Model (CLM) simulations, each with Amazon-like conditions, driven by high frequency forcing data taken from different tropical grid cells of actual vs. emulated SPCAM aqua-planet data. Relative to CLM's conventional coupled behavior (orange lines) these integrations drift to an unusual attractor, which can be understood by the unusual aqua-planet surface state (e.g. high wind speeds from a frictionless surface). Despite this idealization, the key point is that the CLM drifts to the same new attractor regardless of whether the emulated surface inputs or the actual surface inputs are used to drive it, including details of multiple nonlinear cycles that we have traced to threshold physics associated with wildfire and carbon cycle feedbacks interior to CLM's biogeochemistry modules. These similar trajectories, despite the nonlinearities inherent to CLM physics, are strong evidence against the null hypothesis. This supports the idea of trying DNN convection emulators like this in fully interactive real-geography simulations, if they can be adapted to produce all necessary output fields, including separately tracking snow vs. liquid precipitation as well as separating diffuse vs. direct downwelling solar radiation fluxes.

\section{Conclusion \label{sec:Conclusion}}

We find that a feed-forward deep neural network can skillfully emulate the deterministic part of sub-grid scale diabatic heating and moistening tendencies from global superparameterization with the inclusion of land. For the zonal mean, neural network emulated convective tendencies capture over 70\% of the actual variance at the 15-minute sampling scale and over 90\% of the actual variance at the daily-mean sampling scale throughout most of the mid-to upper troposphere. On regional scales, heating skill is best at low altitudes over land, and at mid-levels over extratropical oceans -- both regions where we expect convection to be deterministically set by the large-scale thermodynamic state. On diurnal timescales, convective responses to solar heating are emulated correctly, including land-sea contrast and vertical structure. Full temporal and spatial spectral analyses reveal no obvious ``mode-specifity'' to what is vs. isn't emulated other than imperfections in the goodness of fit on small spatial (less than $10^3$ km in both the zonal and meridional directions) and temporal (less than 3 hours) scales (Figures~\ref{fig:spectra} and~\ref{fig:space_spectra}). A Pearson Correlation Coefficient of 0.50 between DNN skill and autocorrelation statistics suggests these errors are highest in stochastic regions where the deterministic component of diabatic tendencies is weaker such as the tropical, marine boundary layer, and the mid-to-upper troposphere over convective land regions. But on longer timescales, particularly where there are distinct, deterministic patterns of atmospheric variation like the diurnal heating of the continents or baroclinic Rossby wave disturbances along mid latitude storm tracks, our DNN effectively emulates superparameterized diabatic processes. We find the highest $R^{2}$ coefficient of determination values (typically greater than 0.9) for daily and zonal-mean predictions especially compelling (Figure~\ref{fig:15min_vs_daily_zonalmean}c and d). Despite issues in precipitation emulation, our DNN captures much of the marine PDF of precipitation, though it has an unexpected drizzle bias over land that can be partially reduced via a positive constraint on precipitation (Figure~\ref{fig:diurnal_phase_map}). Despite these imperfections, precipitation statistics produced by the DNN are superior to conventional parameterization.

The accuracy achieved by our neural network suggests that feed-forward DNNs may still be the best way to create next generation, hybrid climate emulators. Skip-connections in conjunction with convolutions would seem to have possible advantages in allowing multi-scale structures to be simultaneously prioritized in the loss function~\cite{hanetal2020}. But our DNN achieves similar (Figures~\ref{fig:time_standard_deviation}) to superior (Figure~\ref{fig:Mean_Error}b vs. e), skill compared to the more sophisticated Convolutional (in the vertical direction) Neural Networks and Resnets trained recently on similar data in~\citeA{hanetal2020}. This would suggest that model architecture choices like skip connections and 1D convolutional layers are not critical to achieving a good fit for a neural network in the emulation of convection. 

More broadly, these results also speak to an ongoing question of what sets the ``parameterizability'' of deep convection, which can be inferred from the success of machine learning methods trained on superparameterized simulations (recognizing that despite their constraints SP includes nontraditional degrees of freedom like convective memory and organization). Our findings suggest that convective memory may not be essential ~\cite{hanetal2020, Colin_2019}, at least for feed-forward DNNs. That is to say, our feed-forward DNNs did not require memory from previous timesteps in converging on skillful fits to convective tendencies -- predictions independent of space and time may be the better way forward to achieve successful moist convection emulation. We find that our DNN, with no memory used in training, preferentially fits the atmospheric modes of variation where convective memory would be most helpful (diurnal cycle, onset of afternoon deep convection/heavy precipitation, synoptic storms). Our issues in DNN emulation are greatest over regions where the controlling signals happen at the shortest temporal (or spatial) scales -- especially in the tropical, marine boundary layer. These are exactly the places convective memory would be least helpful. 

Looking ahead, we believe a feed-forward deep neural network, powered by automated hyperparameter tuning as well as physical constraints, may be the most realistic way for ML to emulate superparameterized moist convection in a realistic atmosphere with real-geography boundary conditions. This is also a more direct way to achieve two way coupling with a host climate model since feed-forward DNNs can be rapidly deployed today as prognostic Fortran hybrid models thanks to new automated software~\cite{ott2020fortran}. More broadly, for general deep learning applications, we believe our experience sheds light on the importance of incorporating physical science knowledge while exploiting machine learning methods when designing appropriate neural networks to tackle problems such as moist convection emulation. We have found like many others~\cite{tomsetal2020} that while each of these design choices show notable improvements to emulation performance on their own, both are likely needed in conjuncture to utilize DNNs to their fullest potential.

Though not our primary focus, our findings also point to some of the challenges ahead in neural network emulation of the stochastic component of convection. To replace CRMs in a convection simulation, deep neural networks will likely eventually need to fit not just deterministic but also stochastic parameterizations, which are crucial to error and bias reduction~\cite{krasnopolski_2013, Palmer}. Even in superparameterized climate simulations such stochastic effects, while not critical to mean climate, have been linked to some important regional precipitation extremes~\cite{Randall_2019_Ensemble}. Our results indicate there are high variance modes of moist convection that will be difficult for any feed-forward DNN to represent perfectly without a faster time step interval in training data or stochastic, generative modeling. 

Meanwhile, a next step for the specific case of emulating superparameterization should be an online test of the performance of our trained DNN in prognostic mode to determine if the neural network is skilled enough to produce physically plausible outputs from the coupled run. In this limit, the secondary effects of stochasticity noted by~\cite{Randall_2019_Ensemble} argue deterministic DNNs are appropriate. Although coupling emulated atmospheres to prognostic land models is mostly an unexplored frontier, we are optimistic based on our first pilot tests that it is readily approachable; imperfections of the fit do not break standalone land model simulations. But carrying this forward into fully prognostic coupled tests will require significant work, such as expanding this prototype DNN's output vector to include additional variables needed to allow fully interactive land model coupling, and associated tuning. Even if this next step proves successful, feed-forward DNNs should not be thought of as a panacea for all flaws in climate models -- they cannot in their present application resolve biases induced by imperfect microphysics parameterization and the resulting errors in associated turbulence and cloud-radiative effects produced by superparameterized models. However, neural networks do still have broad use for sidestepping the computational bottlenecks that currently limit the global modeling community's ability to approach eddy-resolving scales. We remain excited about that potential, especially given our findings here that such approaches can be made remarkably skillful beyond aqua-planets, at least in tests of offline hold-out test skill.

\acknowledgments
GM acknowledges support from a NSF graduate fellowship under grant 1633631 and thanks the MAPS Program. GM, MP and TB acknowledge additional funding from NSF grants OAC-1835863 and AGS-1734164. The work of JO and PB in part supported by NSF NRT grant 1633631 to PB. PG thanks NSF OAC-1835863 and REC synergy grant USMILE. The authors thank Stephan Rasp for the initial Github repository cloned for this work and as well as his guidance. We also are grateful to Stephan Mandt and Ruihan Yang for helpful conversations and ideas that advanced this work. The authors thank Chris Terai and Liran Peng for assistance with the simulation of SPCAM and CAM data. Additionally, we want to thank our two anonymous reviewers for their thoughtful feedback as well as JAMES Editor-in-Chief Robert Pincus. Computational resources were provided by the Extreme Science and Engineering Discovery Environment supported by NSF grant number ACI-1548562 (charge number TG-ATM190002). Code for preprocessing, training, and post-processing figure generation can be found at \break ~\url{https://doi.org/10.5281/zenodo.4554598}. All compressed data for figures as well as sample training and test simulation datas can be found at \break ~\url{https://doi.org/10.5281/zenodo.4558716} while details on how to recreate the entire simulation are available in the Methods Section.

\bibliography{sources}

\clearpage
\appendix
\section{Performance Comparison with Existing Literature}

\begin{figure}[ht!]
 \hspace{-3cm}
 \includegraphics[width=20cm]{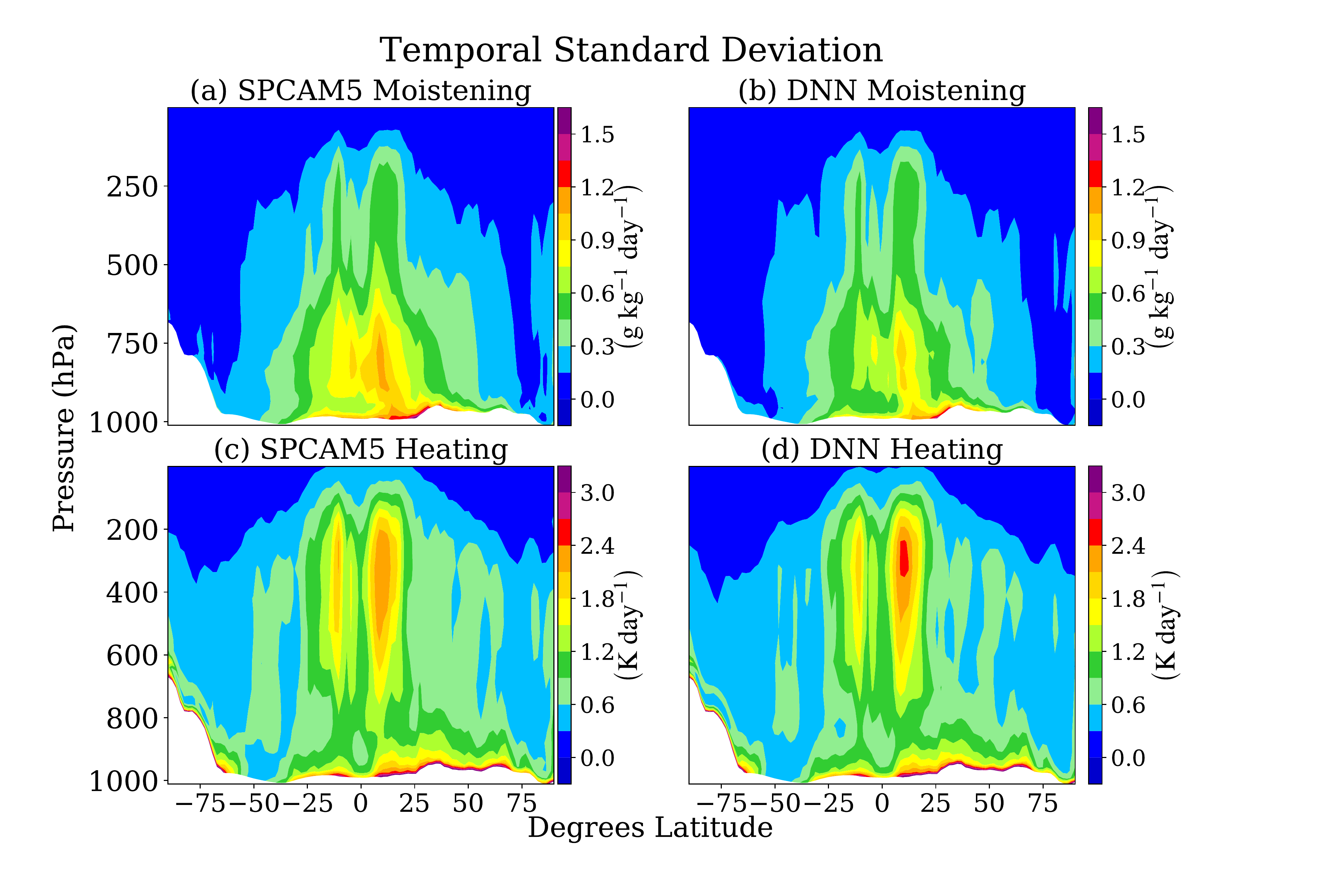}
 \caption{The temporal standard deviation of annual heating and moistening tendencies. Units converted to $(K / day)$ and $(g / kg /day)$ respectively to contrast with the performance of a Resnet~\cite{hanetal2020}.}
 \label{fig:time_standard_deviation}
\end{figure}

Figure~\ref{fig:time_standard_deviation} shows the extent to which the variations of the atmosphere, particularly those driven by deep convection and latent heating can be captured by a feed-forward DNN with minimal under-prediction. The spatio-temporal patterns are replicated over the annual data.

\begin{figure}[ht!]
\centering
\includegraphics[width=\textwidth]{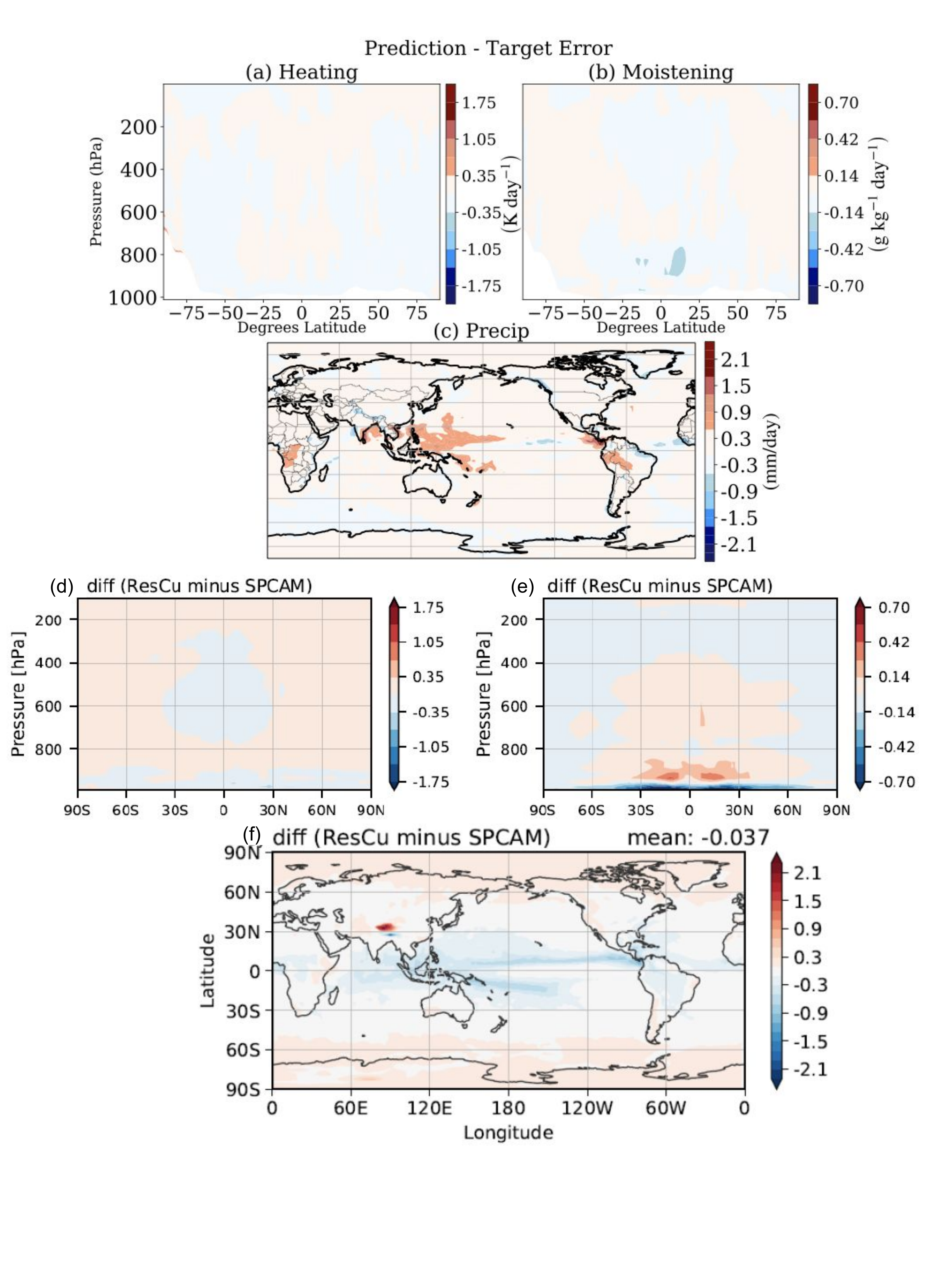}
\caption{The difference between annual target SPCAM5 data and the DNN predictions for heating tendency $(K / day)$, moistening tendency $(g / kg /day)$ and precipitation $(mm / day)$. The 3 panels on the bottom have been taken from~\cite{hanetal2020} to provide direct comparisons between the performance of our DNN and the~\citeA{hanetal2020} Resnet on full complexity, real-geography simulation data.}
\label{fig:Mean_Error}
\end{figure}

Overall, when looking at the annual mean, our DNN preforms well globally. These are some imperfections with emulation of intense tropical precipitation, but the heating and moistening tendency predictions are very close to the target data. In particular, Figure~\ref{fig:Mean_Error} shows modes of variation in the planetary boundary layer can be captured by our DNN. The DNN seems to fit at least as well as the Resnet throughout the latitude-pressure cross section, and perhaps marginally better the boundary layer when moisture variations are examined (Figure~\ref{fig:Mean_Error} b vs. e).

\end{document}


%
%


\title{Supporting Information for Assessing the Potential of Deep Learning for Emulating Cloud Superparameterization in 
Climate Models with Real-Geography Boundary Conditions}
%
%

%
%



\authors{Griffin Mooers\affil{1,2}, Michael Pritchard\affil{1}, Tom Beucler\affil{1}, Jordan Ott\affil{2,1}, Galen Yacalis\affil{4}, Pierre Baldi\affil{2}, Pierre Gentine\affil{3}}

\affiliation{1}{Department of Earth System Science, University of California at Irvine, CA, USA}
\affiliation{2}{Department of Computer Science, University of California at Irvine, CA, USA}
\affiliation{3}{Department of Earth and Environmental Engineering, Columbia University, New York, NY, USA}
\affiliation{4}{Department of Mathematics, University of California at Irvine, CA, USA}

%
%

%

\begin{article}

%
%

\noindent\textbf{Contents of this file}
\begin{enumerate}
\item Figure S1
\item Movies S1 to S2
\item Tables S1 to S4
\end{enumerate}

\noindent\textbf{Introduction}
This supporting information section provides an expanded version of Table 4 for additional context on neural network performance and comparison between the baseline neural network, the manually tuned neural network, and the formally tuned (Sherpa) neural network. We also include an expanded version of Figure 10 to contrast the performances between constrained neural networks looking at the totality of the precipitation PDF instead of just the hour of maximum precipitation view offered in Figure 9. We have also embedded public links for several animations showing neural network emulation of convective tendencies and precipitation in the tropics.

%

\begin{figure}[ht!]
\centering
\includegraphics[width=\textwidth]{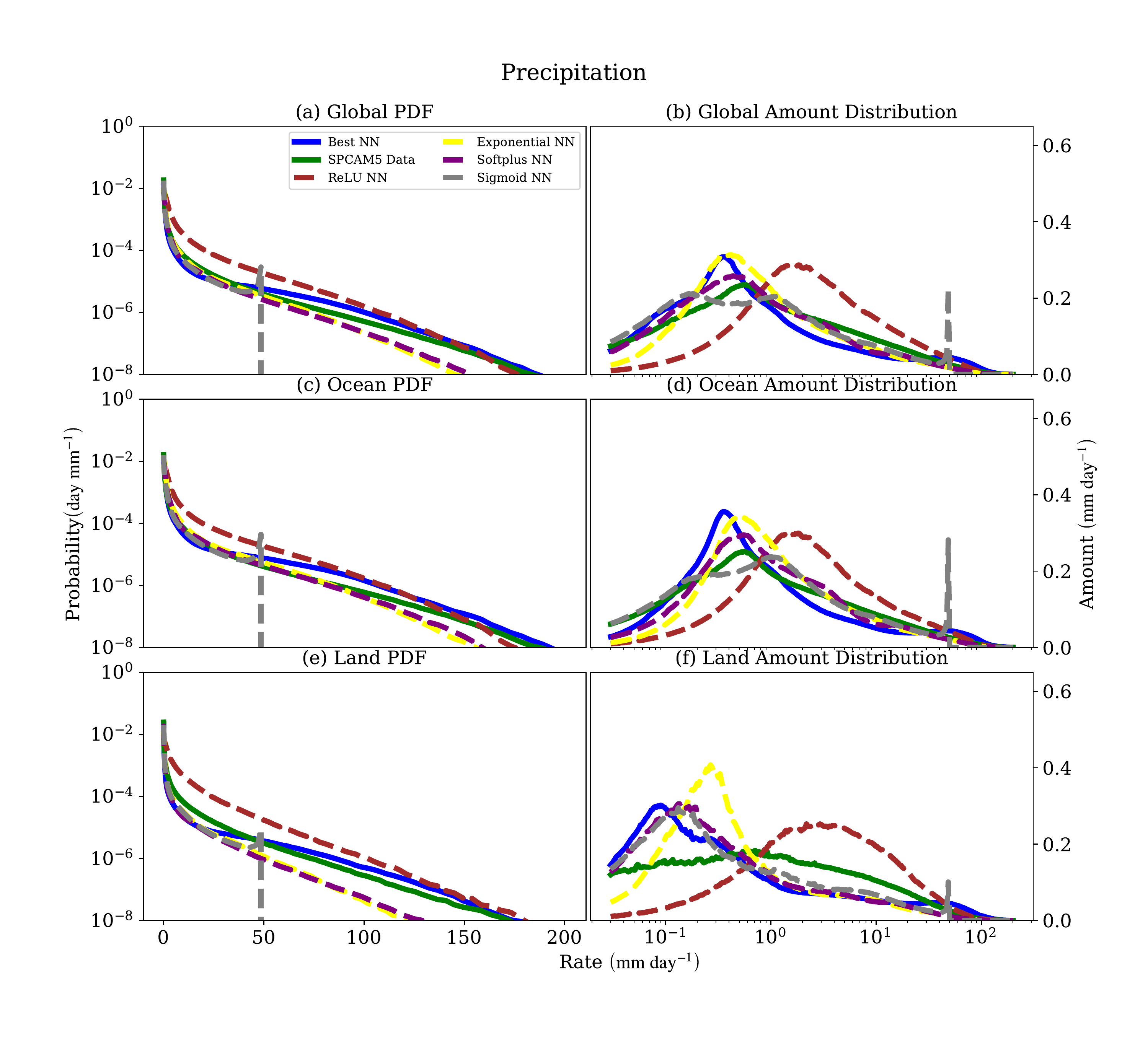}
\caption{\label{fig:Precip_Dist_SI} An extension of Figure 10, but this time contrasting four constrained neural networks (dashed lines) against the SPCAM5 target data (green line) and the Sherpa NN (blue line). The Probability Density Function across the range of simulated precipitation rates (a, c, e) and the amount distribution (b, d, f) of precipitation in which the probability density function is multiplied by the bin-averaged values of precipitation.}
\end{figure}

\begin{table}
 \centering
 \begin{tabular}{|c|c|c|c|c|c|c|c|}
 \hline
  Label & Training Data & Region & Variable & timestep & 25th & 50th & 75th \\ 
  \hline
  a & real-geog. (Linear) & Land & Heating & 15 min. & -0.96 & -0.06 & 0.25\\
  b & real-geog. (Linear) & Land & Heating & Daily & -3.09 & -1.22 & -0.09\\
  c & real-geog. (Manual) & Land & Heating & 15 min. & -0.93 & -0.06 & 0.35\\
  d & real-geog. (Manual) & Land & Heating & Daily & -2.87 & -1.00 & 0.00\\
  e & real-geog. (Best) & Land & Heating & 15 min. & 0.41 & 0.64 & 0.82\\
  f & real-geog. (Best) & Land & Heating & Daily & 0.42 & 0.71 & 0.85\\
  \hline
 \end{tabular}
 \caption{Statistical breakdown of skill score showing percentiles summarizing skill variability in 3D space, i.e. from a flattened vector of $\mathrm{R^{2}}$ values that were calculated across just the time dimension separately for each longitude, latitude, and pressure level, using raw data at the 15-minute sampling scale or the daily mean sampling scale. This table highlights the three neural networks trained on the SPCAM5 real geography data: A "Linear" baseline model (a-b), a manually tuned neural network (c-d), And our formally tuned Sherpa neural network (e-f). The table depicts convective heating $\mathrm{K/s}$ over continental locations.}
 \label{tab:S1}
\end{table}
\clearpage

\begin{table}
 \centering
 \begin{tabular}{|c|c|c|c|c|c|c|c|}
 \hline
  Label & Training Data & Region & Variable & timestep & 25th & 50th & 75th \\ 
  \hline
  a & real-geog. (Linear) & Land & Moistening & 15 min. & -0.07 & -0.02 & 0.00\\
  b & real-geog. (Linear) & Land & Moistening & Daily & -0.27 & -0.09 & -0.02\\
  c & real-geog. (Manual) & Land & Moistening & 15 min. & -0.07 & -0.02 & 0.00\\
  d & real-geog. (Manual) & Land & Moistening & Daily & -0.27 & -0.09 & -0.02\\
  e & real-geog. (Best) & Land & Moistening & 15 min. & -5.5 & 0.10 & 0.55\\
  f & real-geog. (Best) & Land & Moistening & Daily & -12.9 & 0.14 & 0.76\\
  \hline
 \end{tabular}
 \caption{Statistical breakdown of skill score showing percentiles summarizing skill variability in 3D space, i.e. from a flattened vector of $\mathrm{R^{2}}$ values that were calculated across just the time dimension separately for each longitude, latitude, and pressure level, using raw data at the 15-minute sampling scale or the daily mean sampling scale. This table highlights the three neural networks trained on the SPCAM5 real geography data: A "Linear" baseline model (a-b), a manually tuned neural network (c-d), And our formally tuned Sherpa neural network (e-f). The table depicts convective moistening $\mathrm{kg/kg/s}$ over continental locations.}
 \label{tab:S2}
\end{table}
\clearpage


\begin{table}
 \centering
 \begin{tabular}{|c|c|c|c|c|c|c|c|}
 \hline
  Label & Training Data & Region & Variable & timestep & 25th & 50th & 75th \\ 
  \hline
  a & aqua-planet & Ocean & Heating & 15 min. & 0.05 & 0.27 & 0.55\\
  b & aqua-planet & Ocean & Heating & Daily & -0.41 & 0.24 & 0.59\\
  c & real-geog. (Linear) & Ocean & Heating & 15 min. & -0.30 & -0.01 & 0.21\\
  d & real-geog. (Linear) & Ocean & Heating & Daily & -2.26 & -0.58 & 0.03\\
  e & real-geog. (Manual) & Ocean & Heating & 15 min. & -0.26 & 0.00 & 0.31\\
  f & real-geog. (Manual) & Ocean & Heating & Daily & -1.82 & -0.33 & 0.32\\
  g & real-geog. (Best) & Ocean & Heating & 15 min. & 0.28 & 0.54 & 0.76\\
  h & real-geog. (Best) & Ocean & Heating & Daily & 0.30 & 0.66 & 0.85\\
  \hline
 \end{tabular}
 \caption{Statistical breakdown of skill score showing percentiles summarizing skill variability in 3D space, i.e. from a flattened vector of $\mathrm{R^{2}}$ values that were calculated across just the time dimension separately for each longitude, latitude, and pressure level, using raw data at the 15-minute sampling scale or the daily mean sampling scale. This cumulative table compares results from a neural network trained on SPCAM3 aqua-planet data (a-b). It also highlights three neural networks trained on the SPCAM5 real geography data: A "Linear" baseline model (c-d), a manually tuned neural network (e-f), and our formally tuned Sherpa neural network (g - h). The table depicts convective heating $\mathrm{K/s}$ over marine locations.}
 \label{tab:S3}
\end{table}
\clearpage


\begin{table}
 \centering
 \begin{tabular}{|c|c|c|c|c|c|c|c|}
 \hline
  Label & Training Data & Region & Variable & timestep & 25th & 50th & 75th \\ 
  \hline
  a & aqua-planet & Ocean & Moistening & 15 min. & -4e6 & -0.40 & 0.23\\
  b & aqua-planet & Ocean & Moistening & Daily & -4e5 & -2.19 & 0.40\\
  c & real-geog. (Linear) & Ocean & Moistening & 15 min. & -0.06 & -0.01 & 0.48\\
  d & real-geog. (Linear) & Ocean & Moistening & Daily & -0.22 & -0.06 & 0.00\\
  e & real-geog. (Manual) & Ocean & Moistening & 15 min. & -0.05 & 0.00 & 0.04\\
  f & real-geog. (Manual) & Ocean & Moistening & Daily & -0.22 & -0.06 & 0.00\\
  g & real-geog. (Best) & Ocean & Moistening & 15 min. & -3.45 & 0.16 & 0.48\\
  h & real-geog. (Best) & Ocean & Moistening & Daily & -11.0 & 0.49 & 0.49\\
  \hline
 \end{tabular}
 \caption{Statistical breakdown of skill score showing percentiles summarizing skill variability in 3D space, i.e. from a flattened vector of $\mathrm{R^{2}}$ values that were calculated across just the time dimension separately for each longitude, latitude, and pressure level, using raw data at the 15-minute sampling scale or the daily mean sampling scale. This cumulative table compares results from a neural network trained on SPCAM3 aqua-planet data (a-b). It also highlights three neural networks trained on the SPCAM5 real geography data: A "linear" baseline model (c-d), a manually tuned neural network (e-f), and our formally tuned Sherpa neural network (g - h).The table depicts convective moistening $\mathrm{kg/kg/s}$ over marine locations.}
 \label{tab:S4}
\end{table}
\clearpage

\noindent\textbf{Movie S1.} 
An animation of convective heating and moistening tendencies between CAM5 data, SPCAM5 data, and the Sherpa tuned "Best" neural network. This video contains every other 15 minute timestep for 14 days in July from 35S-35N. The data have all been converted to $\mathrm{W/m^{2}}$. The complete animation can be accessed at \break ~\url{https://drive.google.com/file/d/17MdO7Lb7DusakuT_2WcqW7iakVYUr7hh/view}

\noindent\textbf{Movie S2.}

An animation of precipitation from CAM5 data, SPCAM5 data, and the Sherpa tuned "Best" neural network. This video contains every other 15 minute timestep for 14 days in July from 35S-35N. The data have all been converted to $\mathrm{mm/day}$ and the data is shown on a log scale. The complete animation can be accessed at \break ~\url{https://drive.google.com/file/d/1jLccgEBkeIK-ciCvoKKtt0JOdlNPcvRl/view}
\end{article}
\clearpage